\documentclass[traditabstract]{aa} % for the abstract without structuration 
\usepackage{graphicx}
\usepackage{color}
\usepackage{txfonts}
\usepackage{natbib}
\usepackage{psfrag}
\bibpunct{(}{)}{;}{a}{}{,} % to follow the A&A style
\newcommand{\bea}	{\begin{array}}
\newcommand{\eea}	{\end{array}}
\newcommand{\beq}	{\begin{equation}}
\newcommand{\eeq}	{\end{equation}}
\newcommand{\ben}	{\begin{eqnarray}}
\newcommand{\een}	{\end{eqnarray}}
\newcommand{\bsq}	{\begin{mathletters}}
\newcommand{\esq}	{\end{mathletters}}

\newcommand{\Ro}	{R_0}

\newcommand{\vc}	{v_{\rm c}}
\newcommand{\vo}	{v_0}

\newcommand{\vphi}	{v_{\phi}}
\newcommand{\vr}	{v_{R}}
\newcommand{\volrhelio}	{V_{\rm\scriptscriptstyle OLR}}
\newcommand{\volr}	{v_{\phi, \rm\scriptscriptstyle OLR}}
\newcommand{\uolr}	{v_{R, \rm\scriptscriptstyle OLR}}
\newcommand{\volrrot}	{v_{\phi, \rm\scriptscriptstyle OLR}^\theta}
\newcommand{\uolrrot}	{v_{R, \rm\scriptscriptstyle OLR}^\theta}
\newcommand{\Ob}	{\Omega_{\rm b}}
\newcommand{\Vs} 	{V_\odot}
\newcommand{\Us} 	{U_\odot}
\newcommand{\Ws} 	{W_\odot}
\newcommand{\pb}	{\phi_{\rm b}}
\newcommand{\Oo}	{\Omega_0}

\newcommand{\kms}{\rm\,km\,s^{-1}}
\newcommand{\kmskpc}{{\rm\,km\,s^{-1}{kpc}^{-1}}}
\newcommand{\pc}{\rm\,pc}
\newcommand{\kpc}{{\rm\,kpc}}

\newcommand{\Gyr}{{\rm\,Gyr}}

\def\aap{A\&A}

\def\2s{2-$\sigma$}
\def\3s{3-$\sigma$}

\begin{document}
   \title{Constraints on the Galactic Bar from the Hercules stream as traced with RAVE across the Galaxy}% 
          %Tracing the Hercules stream across the Galaxy \\to constrain the Galactic bar}%change
%: \\from the Solar neighbourhood to the district
%   \subtitle{I. Solar neighbourhood}

   \author{T. Antoja\inst{1}
\and
A. Helmi\inst{1}
\and
W. Dehnen\inst{2}
\and
O. Bienaym\'e\inst{3}
\and
J. Bland-Hawthorn\inst{4}
\and
B. Famaey\inst{3}
\and
K. Freeman\inst{5}
\and
B. K. Gibson\inst{6}
\and
G. Gilmore\inst{7}
\and
E.~K.\ Grebel\inst{8}
\and
G. Kordopatis\inst{7}
\and
A. Kunder\inst{9}
\and
I. Minchev\inst{9}
\and
U. Munari\inst{10}
\and
J. Navarro\inst{11}
\and
Q. Parker\inst{12,13,14}
\and
W.A. Reid\inst{12,13}
\and
G. Seabroke\inst{15}
\and
A. Siebert\inst{3}
\and
M. Steinmetz\inst{9}
\and
F. Watson\inst{14}
\and
R.F.G. Wyse\inst{16}
\and
T. Zwitter\inst{17,18}
\fnmsep%\thanks{Just to show the usage           of the elements in the author field}
          }

   \institute{Kapteyn Astronomical Institute, University of Groningen, PO Box 800, 9700 AV Groningen, the Netherlands\\
              \email{tantoja@rssd.esa.int}
  \and
University of Leicester, University Road, Leicester LE1 7RH, UK
\and
Universit\'e de Strasbourg, CNRS, Observatoire, 11 rue de l'Universit\'e F-67000 Strasbourg, France
\and
Sydney Institute for Astronomy, School of Physics, University of Sydney, NSW 2006, Australia 
\and
RSAA Australian National University, Mount Stromlo Observatory, Cotter Road, Weston Creek, Canberra, ACT 2611, Australia
\and
 Jeremiah Horrocks Institute, University of Central Lancashire, Preston, PR1 2HE, United Kingdom
\and
Institute of Astronomy, University of Cambridge, Madingley Road, Cambridge, CB3 0HA, UK
\and
Astronomisches Rechen-Institut, Zentrum f\"ur Astronomie der Universit\"at Heidelberg, M\"onchhofstr.\ 12--14, 69120 Heidelberg, Germany 
\and
Leibniz-Institut f\"ur Astrophysik Potsdam (AIP), An der Sternwarte 16, D - 14482, Potsdam, Germany
\and
INAF Osservatorio Astronomico di Padova, 36012 Asiago (VI), Italy 
\and
Senior CIfAR Fellow. University of Victoria, Victoria BC Canada V8P 5C2
\and
Department of Physics and Astronomy, Macquarie University, Sydney, NSW, 2109, Australia
\and
Research Centre for Astronomy, Astrophysics and Astrophotonics, Macquarie University, Sydney, NSW 2109 Australia
\and
Australian Astronomical Observatory, PO Box 915, North Ryde, NSW 1670, Australia
\and
Mullard Space Science Laboratory, University College London, Holmbury St Mary, Dorking, RH5 6NT, UK
\and
Johns Hopkins University, Homewood Campus, 3400 N Charles Street, Baltimore, MD 21218, USA
\and
University of Ljubljana, Faculty of Mathematics and Physics, Jadranska 19, 1000 Ljubljana, Slovenia
\and
Center of excellence Space-SI, Askerceva 19, 1000 Ljubljana, Slovenia          }

   \date{Received XX; accepted XX}

% \abstract{}{}{}{}{} 
  \abstract{
Non-axisymmetries in the Galactic potential (spiral arms and bar) induce kinematic groups such as the Hercules stream. Assuming that Hercules is caused by the effects of the Outer Lindblad Resonance of the Galactic bar, we model analytically its properties as a function of position in the Galaxy and its dependence on the bar's pattern speed and orientation. Using data from the RAVE survey we find that the azimuthal velocity of the Hercules structure decreases as a function of Galactocentric radius, in a manner consistent with our analytical model. This allows us to obtain new estimates of 
the parameters of the Milky Way's bar. The combined likelihood function of the bar's pattern speed and angle has its maximum for a pattern speed of $\Ob=(1.89 \pm 0.08) \times \Oo$, where $\Oo$ is the local circular frequency. 
 Assuming a Solar radius of $8.05\kpc$ and a local circular velocity of $238 \kms$, this corresponds to $\Ob= 
56 \pm 2 \kmskpc$. %%Additional systematic errors could make the reported errors larger.
On the other hand, the bar's orientation $\pb$ cannot be constrained with the available data. %but that a high value seems to be preferred.  
In fact, the likelihood function shows that a tight correlation exists between the pattern speed and the orientation, implying
that a better description of our best fit results is given by the linear relation $\Ob/\Oo=1.91 + 0.0044\left({\pb}(\deg)-48\right)$, with standard deviation of $0.02$. 
%%obtained through the covariance matrix. 
For example, for an angle of $\pb=30\deg$ the pattern speed is $54.0\pm0.5\kmskpc$. 
% tight correlation between Omega and phi and the tighter constraint on the appropriate combination. 
 These results are not very sensitive to the other Galactic parameters such as the circular velocity curve or the peculiar motion of the Sun, and are robust to biases in distance.
}


   \maketitle

%________________________________________________________________

\section{Introduction}\label{intro}

The existence of a bar in our Galaxy is supported by a variety of
studies using data from HI 21cm and CO emission, star counts in the
Galactic Centre (GC), IR observations from DIRBE (Diffuse InfraRed Background Experiment) on COBE (COsmic Background Explorer) and GLIMPSE (Galactic Legacy Infrared Mid-Plane Survey Extraordinaire) with Spitzer, or microlensing
surveys (see \citealt{Gerhard02} for a review). However, previous
research has revealed inconsistent results regarding the characteristics of the bar. 
For example, estimates of its pattern speed range from $40$ to $65\kmskpc$
\citep{Gerhard11} 
while the estimates of its orientation with respect to the Sun range from 
$10\deg$ \citep{LopezCorredoira00,Robin12} to $45\deg$
\citep{Hammersley00,Benjamin05}. 
The presence of a secondary bar in our Galaxy is also currently under debate
\citep{MartinezValpuesta11,RomeroGomez11}.

\citet{Kalnajs91} presented an indirect method to measure 
the bar properties based on the location of kinematic structures in
the Solar neighbourhood. He related the velocities of the 
Hyades and Sirius moving groups to the two types of orbits expected 
around the Outer Lindblad Resonance (OLR), and in this way
constrained the bar's pattern speed and its orientation. 

Many more substructures in the local velocity distribution were
unveiled by the ESA's astrometric mission Hipparcos
\citep[e.g.][]{Dehnen98,deZeeuw99,Chereul99}. Most of these groups were
initially thought to be remnants of disrupted clusters (Eggen
1996). However, there is evidence of a 
large scatter in age and metallicity in some of them
\citep{Raboud98,Dehnen98,Skuljan99,Famaey05,Bobylev07,Antoja08}.
Therefore, it is likely that these substructures formed as a response to
the non-axisymmetries of the gravitational potential rather than being groups of stars
of a common origin.

Several studies after \citet{Kalnajs91} have attempted to use
these local velocity groups to better constrain the properties of the Galactic bar
\citep[e.g.][hereafter D00]{Dehnen00}, and also of the spiral structure
\citep[e.g.][]{Quillen05}. However, \citet{Antoja09,Antoja11} have
shown that the groups
detected in the Solar vicinity can be reproduced by models with
different parameters, including bar or/and spiral structure, highlighting
that local estimates are subject to degeneracies.

The simulations of \citet{Antoja11,Antoja09} as well as those of e.g.\
\citet{Quillen11} have shown that the groups' kinematics
change across the disc. Recently, using action-angle modelling
\citet{McMillan13} showed how the local Hyades stream can be due to
the effects of different resonances such as the Inner Lindblad Resonance or OLR of a
non-axisymmetric pattern in the disc. But these models predict
differences in the stream kinematics throughout the disc.

With the advent of data from new surveys such as RAVE (RAdial Velocity
Experiment, \citealt{Steinmetz06}) the detection of kinematic
groups is no longer limited to the Solar vicinity. The first example
was given by \citet{Antoja12} (hereafter A12) where wavelet
transform techniques were used to detect kinematic groups beyond
the Solar neighbourhood in the RAVE survey. The sampled volume allowed
the demonstration that some local groups can be traced at least up to $1\kpc$ away
from the Sun in certain directions and that their
velocities change with distance. These discoveries point toward the 
exciting possibility of using observed velocity distributions in a number of regions of the Galaxy to break degeneracies and eventually constrain the properties of the spiral arms and the bar.

A12 showed that Hercules, a local group of stars moving outwards in
the disc and lagging the Local Standard of Rest, has a larger
azimuthal velocity inside the Solar circle and a smaller one outside.
Here we quantify this trend in more detail with the new RAVE DR4, showing that it is consistent with the effects of the bar's OLR, and we use it to constrain the properties of the Galactic bar.

In Sect.~\ref{in} we review the properties of the local Hercules stream and its relation with the effects of the bar's OLR. Also by extending the modelling work of D00, we derive an analytic expression for the variation of the azimuthal velocity of Hercules as a function of Galactocentric radius for different bar properties. We then use 
simulations of a 
barred disc to test this model and the recovery of the simulation's parameters (Sect.~\ref{test}). In Sect.~\ref{application} we measure the observed azimuthal velocity of Hercules as a function of radius for RAVE stars. We finally compare these measurements with the predictions of the effects of the bar's OLR and we derive the best fit parameters of the bar (Sect.~\ref{results}). 
Section \ref{conclusions} contains a final discussion and conclusions.

%__________________________________________________________________

\section{The Hercules stream}\label{in}

\subsection{The local Hercules stream and the OLR}\label{intro2}

 \begin{figure}
   \centering
   \includegraphics[width=0.35\textwidth]{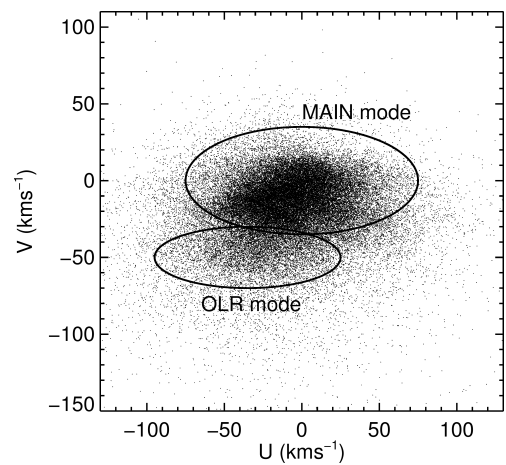}
 \caption{Heliocentric velocities of the Solar Neighbourhood from the RAVE local sample of A12. We have marked the location of the most important kinematic modes,
named here  OLR (Hercules stream) and MAIN modes}
         \label{RAVEvsMODEL}
   \end{figure}

The Hercules moving group (also refereed to as the {\it{U-anomaly}}) was initially identified %in samples of late type stars %check
by \citet{Eggen58} as a group of 22 stars with velocities similar to the high velocity star $\xi$ Herculis. Also \citet{Blaauw70} noticed an excess of negative $U$ velocities (with $U$ directed to the Galactic centre) for stars with $V\sim - 50\kms$, that is a stream of stars with eccentric orbits with a mean outward radial motion. The Hercules stream is also evident in the Solar Neighbourhood velocity distribution of RAVE stars (Fig.~\ref{RAVEvsMODEL}). By using photometric data for a sub-sample of Hipparcos stars \citet{Raboud98} showed that the Hercules metallicity distribution covers the whole range observed in the old disc ($-0.6$ dex to $+0.6$ dex) and a heterogeneous distribution of ages between 6 to $10 \Gyr$. Similar conclusions were obtained using colour as a proxy for age \citep{Dehnen98}, with isochrones for giant stars \citep{Famaey05}, with ages from Str\"omgren photometry and spectro-photometric metallicities for F and G dwarfs \citep{Helmi06,Bobylev07,Antoja08}, and also with high-resolution 
abundances for F and G dwarf stars \citep{Bensby07}. 
Although there is some discrepancy regarding the lower age limit of the group, which ranges from $1$ to $6\Gyr$, it is now clear that this group does not originate in a single population or cluster.
% found that some kinematic groups, including Hercules, are present in various  subsamples of Hipparcos stars. In particular, Hercules can be seen in his sample with a maximum age of $\sim2\Gyr$, its presence becoming stronger in samples dominated by old stars. \citet{Bensby07} analysed 
%Several analysis of the ages of Hercules stars from the Geneva-Copenhagen Survey \citep{Nordstrom04} also show wide spread in age but a predominantly old population \citep{Helmi06}, a mimimum age of $1\Gyr$ and mean age around $4$-$6 \Gyr$ \citep{Bobylev10}. However, according to the two latest mentioned studies the shape of this group could vary with age, being more elongated or branch-like shaped for older samples. The dispersion in the [Fe/H] abundance for Hercules stars is also large (around $0.2\rm dex$ in \citealt{Helmi06} and a slightly larger value of $0.27\rm dex$ in \citet{Antoja08} and also compared to other local groups).
%and peaks close to the disc values. 
%Similar results are found in \citet{Skuljan99,Taylor00,Famaey05,Bovy10}.
%The mean [Fe/H] abundance of the structure was found to be of -0.15, slighlty lower than in Helmi -0.13.
%the presence of the same kinematic groups in samples with early and late-type stars \citep{Dehnen98,Skuljan99}, the isochrones for giant stars in these groups \citep{Famaey05}, ages from Stromgren photometry for F and G dwarf \citep{Bobylev2007,Antoja08} and detalied abundances \citep{Raboud98} indicated that the main groups have . 

The first dynamical models for Hercules were presented in D00 and \citet{Fux01}, and were based on the effects of the bar on the local velocity distribution. D00 proposed that Hercules consists of stars that have been scattered by the OLR. In particular, for certain ranges of pattern speeds and orientations of the bar, a group of unstable orbits ($x_1^*(2)$ orbits) divides the velocity distribution into two main groups (bi-modality) separated by a valley (Fig.~\ref{RAVEvsMODEL}, see also Sect.~\ref{test}). One group is approximately centred on the $U$--$V$ velocity plane (MAIN mode) and the other one has a slower rotation, mean outward radial motion and is associated to the Hercules moving group (OLR mode).

   \begin{figure}
   \centering
   \includegraphics[width=0.45\textwidth]{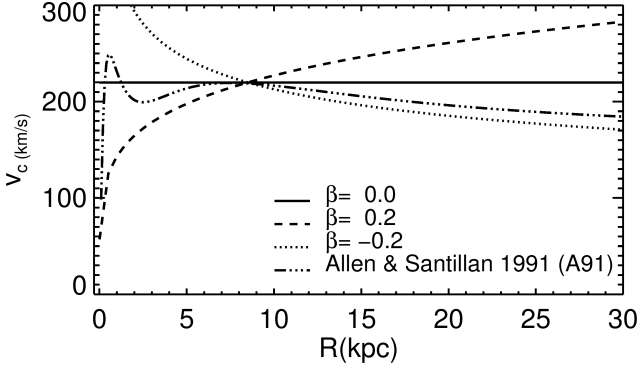}
      \caption{Rotation curves of Eq.~\ref{e:rotcurve} with different values for $\beta$: flat $\beta=0$ (solid line), raising $\beta=0.2$ (dashed line), and decreasing $\beta=-0.2$ (dotted line) rotations curves. The dotted-dashed line shows the rotation curve of the \citet{Allen91} model (A91).}
         \label{rotcurve}
   \end{figure}

D00 simulated the velocity distribution at the Solar position of a barred potential using two-dimensional (in-plane) test particle orbital integrations with the backwards integration technique. The bar model was a quadrupole potential (his Eq.~3) rotating with speed $\Ob$ and orientation angle $\pb$ with respect to the line Sun-GC. He used a simple underlying potential (his Eq.~2b) with a power-law rotation curve of the form:
\beq \label{e:rotcurve}
	\vc = \vo\,(R/\Ro)^\beta,
\eeq
where $\Ro$ denotes the Sun's distance from the GC and $\vo$ the local circular speed. Figure \ref{rotcurve} shows the rotation curves in this model for different values of $\beta$. 

\begin{table}[t]
\caption{Best-fit values for $(a,b,c)$ in Eq.~\ref{e:fit} obtained in D00.}\label{tab:fit}
    \centering
   \begin{tabular}{lccc} 			\\[-2ex] 
      \hline\hline 	\\[-2.5ex]
      $\pb (\deg)$ 	& $a$ & $b$ & $c$ 		\\
      \hline		\\[-2.3ex]
      15	&1.3549	&0.0761	&0.1362	\\
      20	&1.2686	&0.0642	&0.1120	\\
      25	&1.2003	&0.0526	&0.0892	\\
      30	&1.1424	&0.0406	&0.0711	\\
      35	&1.0895	&0.0298	&0.0538	\\
      40	&1.0420	&0.0200	&0.0423	\\
      45	&1.0012	&0.0103	&0.0316	\\
      50	&0.9653	&0.0012	&0.0238	\\	\hline
    \end{tabular}
\end{table}

By considering only these axisymmetric power-law potentials (thus neglecting the effect of the quadrupole bar), whose orbital frequencies can be derived analytically and by dismissing terms of O($v^3/\vo^3$), D00 showed that the stars on unstable resonant orbits exactly on the OLR of the rotating frame form a parabola in velocity space (the valley) described by:
\beq \label{e:volr}
	V +{U^2\over2\vo} \cong 
\vo\tilde\volrhelio \equiv \,{1+\beta\over1-\beta} 
	\left[1-{\Ob/\Oo\over1+\sqrt{(1+\beta)/2}} \right],
\eeq
where $U$ and $V$ are the velocities with respect to the Local Standard of Rest and $\Oo$ is the local circular frequency. This parabola has a maximum at $V=\tilde\volrhelio$ occurring at $U=0$ (a saddle point). However, in his simulations including the quadrupole bar the saddle point between the two modes appears shifted in $U$ and also in $V$ with respect to the analytic estimate given by Eq.~(\ref{e:volr}). Then he found that the $V$-velocity of the saddle-point $\volrhelio$ could be fitted by:
\beq \label{e:fit}
	\volrhelio \approx a\,\tilde\volrhelio - (b + c\,\beta)\,\vo,
\eeq
where the values of $a, b$ and $c$ are reproduced in Table \ref{tab:fit} and depend on the bar's orientation $\pb$. Using Eqs.~\ref{e:volr} and \ref{e:fit} we find:
\beq \label{e:fit2}
	\volrhelio \approx a\, \vo\,{1+\beta\over1-\beta} 
	\left[1-{\Ob/\Oo\over1+\sqrt{(1+\beta)/2}} \right]- (b + c\,\beta)\,\vo,
\eeq
which relates the position of the Hercules saddle point $\volrhelio$ to the pattern speed of the bar, its orientation (through parameters $a$, $b$ and $c$) and the slope and normalisation of the rotation curve. Using the local observed velocity distribution of Hipparcos stars D00 found this saddle point to be at $\volrhelio = (-31 \pm 3) \kms$.

% degeneracy,...
%__________________________________________________________________
\subsection{Analytic model for the Hercules stream across the Galaxy}\label{model}

Our purpose is now to generalise Eq.~\ref{e:fit2} to different Galactocentric radii $R$ (i.e.\ not necessarily the Solar neighbourhood). For this, we replace the quantities describing the Solar neighbourhood by their respective functional forms, that is $\vo$ by $\vo(R/\Ro)^\beta$ and $\Oo$ by $\vo(R/\Ro)^\beta/R$ (Eq.~\ref{e:rotcurve}). In cylindrical Galactocentric coordinates ($\vphi =V+\vo$), Eq.~\ref{e:fit2} becomes:
%\beq \label{e:fit3}
%	\volr \approx a\, \vo\,{1+\beta\over1-\beta} 
%	\left[1-{\Ob/\Oo\over1+\sqrt{(1+\beta)/2}} \right]- (b + c\,\beta-1)\,\vo.
%\eeq
%We may replace the quantities describing the Solar neighbourhood by their respective functional forms, that is $\vo$ by $\vo(R/\Ro)^\beta$ and $\Oo$ by $\vo(R/\Ro)^\beta/R$:
\ben \label{e:fitR}
\volr (R)\approx a\, \vo\,(R/\Ro)^\beta\,{1+\beta\over1-\beta} 
	\left[1-{\Ob R\over\vo(R/\Ro)^\beta}{1\over1+\sqrt{(1+\beta)/2}} \right] \nonumber \\
- (b + c\,\beta-1)\,\vo\,(R/\Ro)^\beta.
\een
In this way, the position of the saddle point between the Hercules and the MAIN mode is a function $\volr= f(\Omega_b,\vo, \beta, \pb, R)$. Now $\pb$ is the angle between the considered region (not necessarily the Solar neighbourhood) and the bar. As explained in D00, a higher (lower) force of the bar creates more (less) pronounced features in the velocity plane but does not influence significantly $\volr$, and, therefore, it does not appear
explicitly in Eqs.~(\ref{e:volr}) and (\ref{e:fit}).

   \begin{figure}
   \centering
   \includegraphics[width=0.45\textwidth]{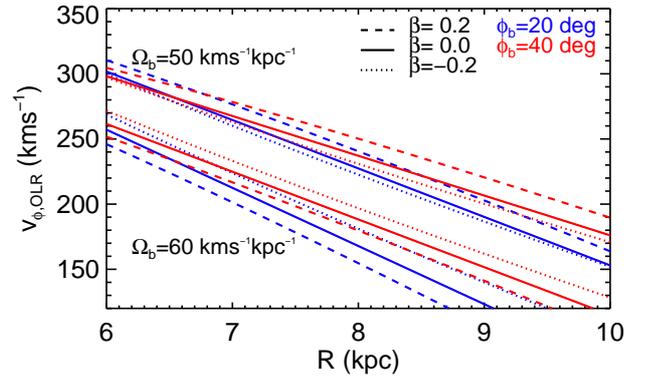}
      \caption{Position of $\volr$ as a function of $R$ for several bar parameters and rotation curves. The normalisation of the rotation curve is here $\vo=238\kms$ and the Solar radius is $\Ro=8.05\kpc$.        }
         \label{theoriccurves}
   \end{figure}

Figure \ref{theoriccurves} shows $\volr$ as a function of Galactocentric radius $R$ for different bar properties and different rotation curves. For this plot we have set a Galactocentric radius of the Sun of $\Ro=8.05\kpc$ and a circular velocity at the Sun of $\vo=238\kms$ following recent results by \citet{Honma12} based on VLBI astrometry of Galactic maser sources. We see that the position of the saddle point decreases with $R$. This is in agreement with what was
reported by A12 for Hercules in the RAVE data. According to the model, the position of the saddle point decreases linearly for $\beta=0$ as: 
\beq
\volr (R)\approx (a-b +1)\,\vo -{a\Ob \over 1+\sqrt{1/2}}R.
\eeq
We see also that for a given angle (shown by lines with the same colour), a higher pattern speed $\Ob$ (lower group of curves) produces a smaller $\volr$ at a given radius compared to lower $\Ob$ (upper group). For a fixed pattern speed, larger bar orientations (red curves) lead to a relation between saddle point $\volr$ and $R$ that has a different slope compared to smaller angles (blue curves). We also notice a slight dependence on the slope of the rotation curve $\beta$: depending on the pattern speed, decreasing (dotted) or increasing (dashed) rotation curves give smaller or larger $\volr$ compared to flat (solid) rotations curves. This dependence is due to the resonances moving closer or farther away from the Sun as the rotation curve is changed.

%__________________________________________________________________
%__________________________________________________________________

\section{Validating the analytical model with simulations}\label{test}

In the previous section we presented an analytical model for the Hercules stream 
which relied on specific assumptions on the potential, namely the shape of the rotation curve, and on the derived frequencies of the orbits. To test the validity of this model, especially Eq.~(\ref{e:fitR}), in this Section we use an independent 
simulation that has been run using a different potential.

%__________________________________________________________________
\subsection{Test particle simulations}\label{test0}

We use a simulation similar to that of \citet{Antoja09}, in which the bar's pattern speed  is $\Ob=47.5\kmskpc$.  
Our simulation uses the same quadrupole bar as in D00 and is also two-dimensional.
 However, our axisymmetric potential is given by \citet{Allen91} (A91), and
composed of a bulge and a flattened disc modelled as Miyamoto-Nagai potentials, and a spherical halo. This axisymmetric model uses a value\footnote{Note that these values are different from the recent ones by \citet{Honma12} that we used for Fig. \ref{theoriccurves} and that we will also use for the RAVE data in Sect. \ref{application}. However, this does not affect our results or conclusions as our analytical formula is general and can be used for any set of parameters.} of $\Ro=8.5 \kpc$ for the Solar radius and a local circular speed of $\vo = 220\kms$.
%The main adopted observational constraints of the model are a value of $\Ro=8.5 \kpc$ for the Solar radius and a rotation curve compatible with several observational studies such as \citet{Fich89}, with a local circular speed of $\vo = 220\kms$.} 
The resulting circular velocity of the model is shown in Fig.~\ref{rotcurve} (dotted-dashed line). This curve 
is fairly different than the power-law models of Eq.~\ref{e:rotcurve} by D00, and presents sections with different slopes and normalisations. The inner peak is due to the presence of the bulge. %For inner radii ($2.5<R<7.5\kpc$) the A91 curve is slightly rising and starts to decrease at $R=8\kpc$. 
%the best fit for the slope $\beta$. For the intermediate radii ($7.5<R<12.\kpc$) and for large radius ($R>10\kpc$) the rotation curve with is slightly decreasing with $\beta\sim-0.094$ and $\beta\sim-0.13$, respectively, having fixed $\vo= 220\kms$. 
This different underlying model does not have the same orbital frequency dependencies used to derive Eq.~\ref{e:volr} and, therefore, allows us to test if the approximations are nonetheless valid for other potentials.

Another important difference of our simulation as compared with D00 is that 
 we use different initial conditions and a different integration scheme. Instead of the backwards integration, we start with $12\cdot10^6$ test particles
with an initial distribution function as discussed in \citet{Hernquist93}. The density follows an exponential disc and the velocity distribution is adopted as a Gaussian with a radial velocity dispersion decaying exponentially with radius, with value of $\sim50\kms$ at the Solar radius. The azimuthal velocity dispersion is related with the radial one through the epicyclic approximation and the asymmetric drift is also taken into account.
The initial conditions generated in this way are not fully consistent with the potential and we expect
these to change in time until reaching stationarity. To avoid these transient effects we first let our initial conditions evolve in the axisymmetric potential for several $\Gyr$ (see \citealt{Monari13} for a discussion). Afterwards, we introduce the bar abruptly in the potential\footnote{ Although we could have introduced the bar slowly in the potential, it has been shown in \citet{Minchev10} that the duration time of the introduction of the bar does not influence significantly the kinematic substructures produced by it. It only changes the time when these effects appear but not the position of the kinematic substructures, nor the position of $\volr$.}
and the final distribution is obtained through forward integration of the orbits for $0.4 \Gyr$ (equivalent to $\sim3$ bar's rotation). 
We consider the particles in a given volume to study the velocity distribution of a particular position in the disc. This is in contrast to D00, whose results correspond to a single position in configuration space.

  \begin{figure}
   \centering
   \includegraphics[width=0.35\textwidth]{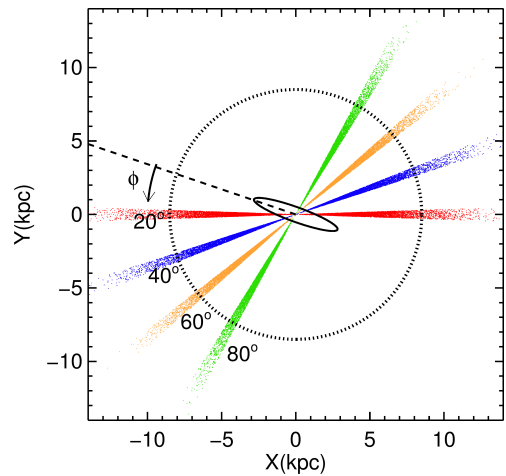}
  \caption{View of the simulated disc with the different bands selected. The bar is indicated as an ellipse with an orientation of $20\deg$ with respect to the Sun (at X=-8.5 and Y=0).}
         \label{xysim}
   \end{figure}

Figure \ref{xysim} shows a sketch of the face-on view of the simulation, with the Sun at $X=-8.5 \kpc$ and $Y=0$, the bar oriented with $\pb=20\deg$ with respect to the Sun, and the Galaxy rotating clockwise. From this simulation we have selected the particles located in 4 different bands with orientations of $20\deg$ (the assumed Sun's position), $40\deg$, $60\deg$ and $80\deg$ with respect to the bar. The width of these bands is $\Delta\phi=4\deg$ and we add the particles in the symmetric bands at respective angles of 
$200\deg$, $220\deg$, $240\deg$ and $260\deg$, which are dynamically equivalent. We now aim to explore the velocity distribution $(\vr, \vphi)$ of these bands as a function of Galactocentric radius. We set $\vr$ positive towards the GC as $U$. We take bins in radius of a width of $\Delta_R=600\pc$ every $600\pc$. The number of stars per bin ranges between 2000 stars for the 
outermost bins to 10000 stars for the innermost ones (Figure \ref{vRsim} top).

\begin{figure*}
   \centering
   \includegraphics[height=0.5\textheight]{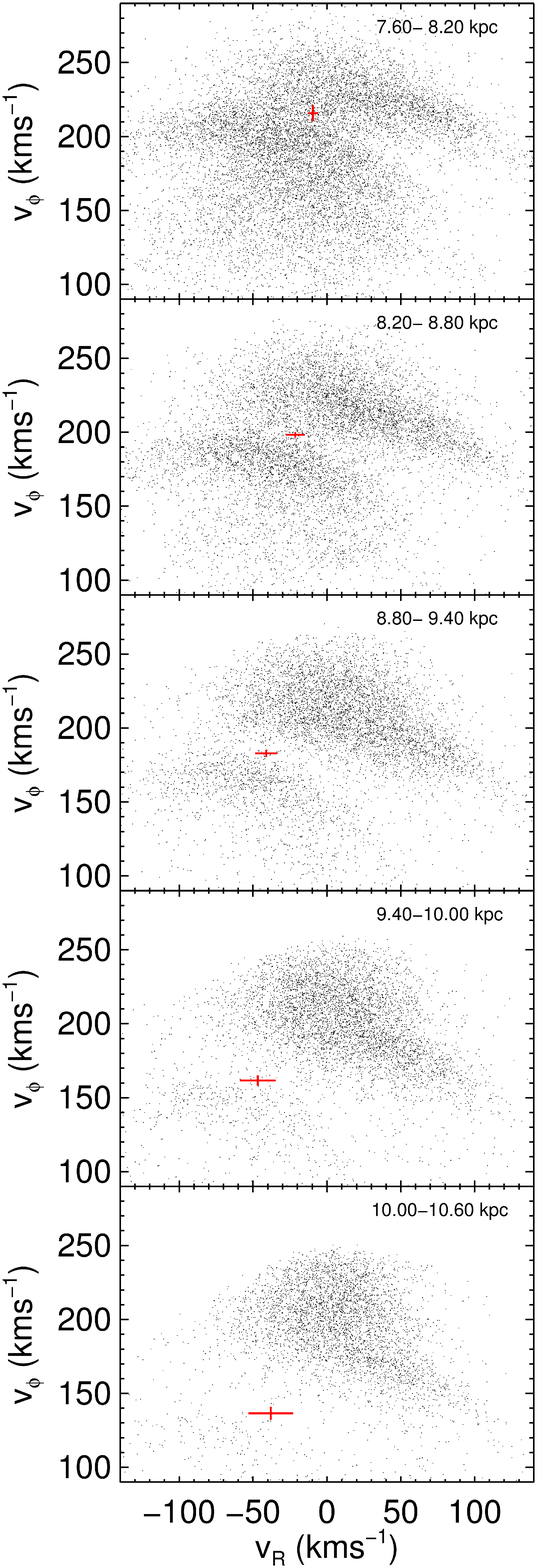}
   \includegraphics[height=0.5\textheight]{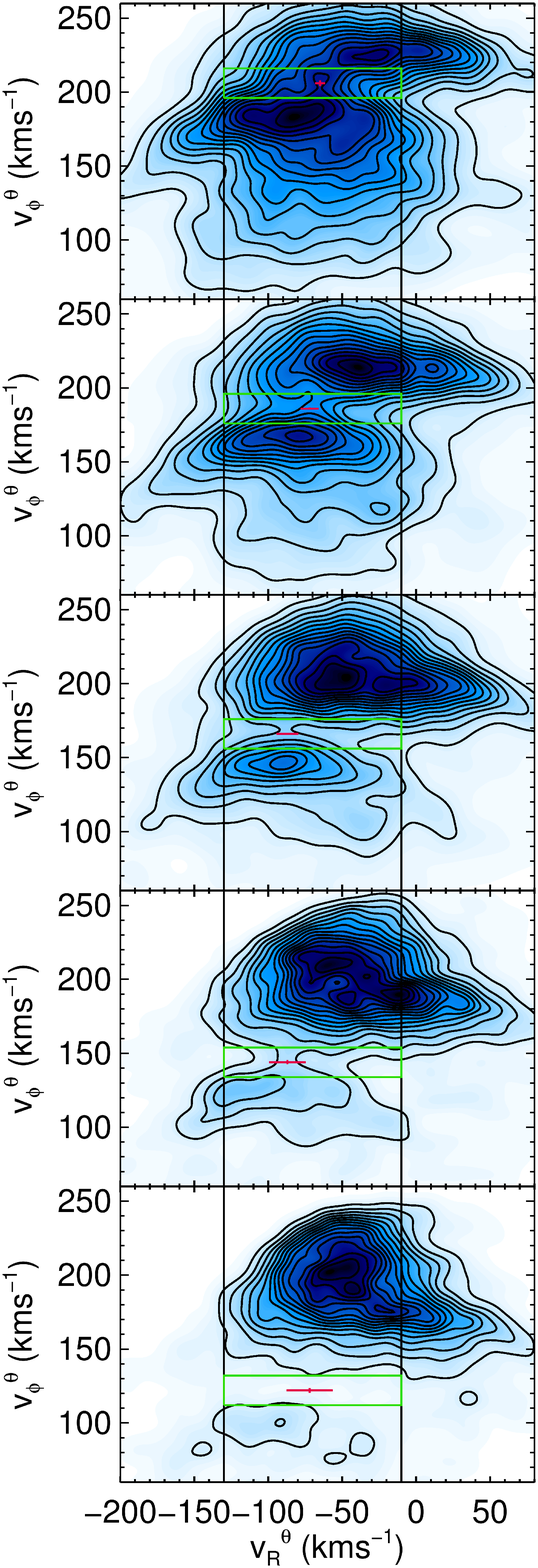}
   \includegraphics[height=0.5\textheight]{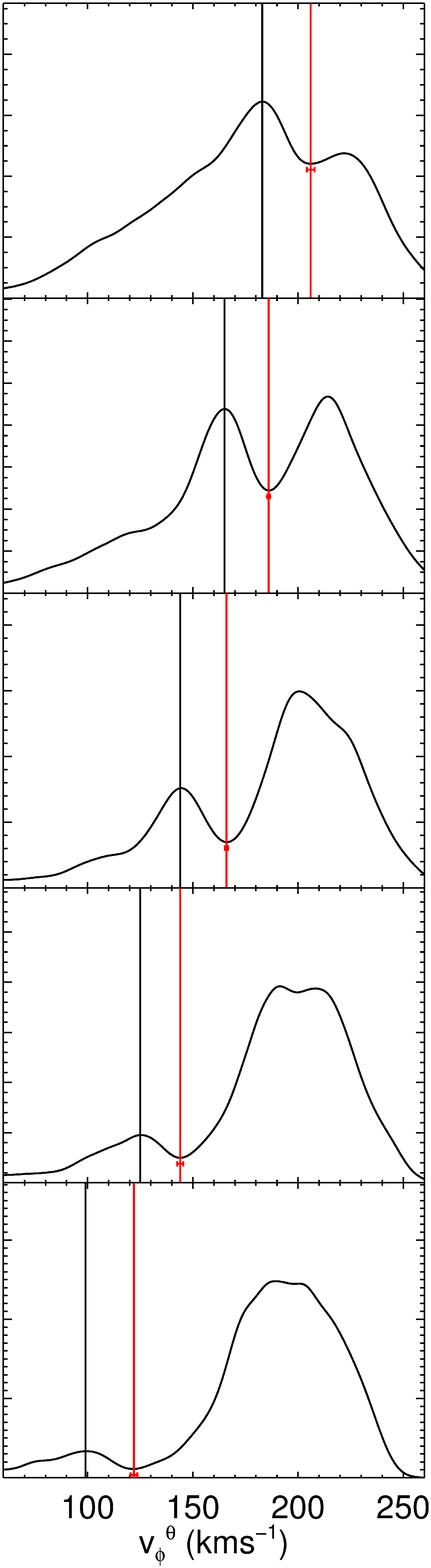}
   \includegraphics[height=0.5\textheight]{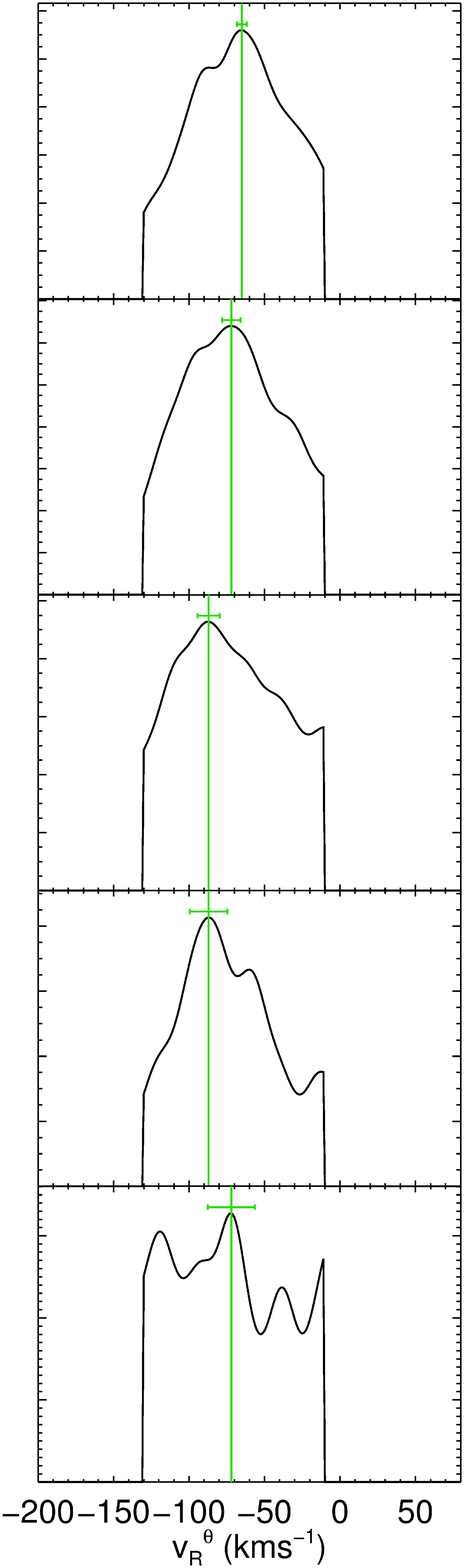}
  \caption{Illustration of the steps followed to locate the saddle point $\volr$ for the band at $40\deg$. 
First column: scatter plot of the velocities in bins in radius as indicated in the top right part of the panels. The red cross in these panels shows our determination of the saddle point. 
Second column:  velocity distribution computed with the kernel adaptive method after a rotation of coordinates by an angle $\theta=15\deg$. 
Third column: distribution of $v_\phi^\theta$ inside the region limited by black vertical lines shown in the second column. The red vertical line and red error bar show the position of the minimum corresponding to the $v_\phi^\theta$ velocity of the saddle point and its error $e_{bst}$. The black vertical line indicates the Hercules peak.
Fourth column:  distribution in $v_R^\theta$ inside the green rectangle shown in the second column. The green vertical line and green error bar shows the position of the maximum which corresponds to the $\uolrrot$ velocity of the saddle point and its error $e_{bst}$.
}
         \label{findvolrsim}
   \end{figure*}

In the first column of Fig.~\ref{findvolrsim} we show these velocity distributions for the band with $40\deg$ orientation (blue band in Fig.~\ref{xysim}). These panels reveal a bimodal distribution, with the structure at lower $\vphi$ and negative $\vr$ being the modelled Hercules group. For this band, we take bins from  $R=7.6\kpc$ to $R=10.6\kpc$ as this is the range for which 
Hercules can be traced. Simple visual inspection shows that, as predicted by our model, the $\vphi$ velocity of Hercules (or equivalently the velocity of the saddle point $\volr$) decreases as a function of $R$. In the next section we show how we measure the velocity of the saddle point $\volr$. 

%__________________________________________________________________

\subsection{Measuring $\volr$ in the simulations}\label{finding0}

We measure the position of the saddle point $\volr$ as illustrated in Fig.~\ref{findvolrsim} for the band at $40\deg$:
\begin{enumerate}
\item \label{i:rotate}We rotate the ($v_\phi$, $v_R$) velocities to align the Hercules structure with the horizontal axis, leading to the new coordinates ($v_R^\theta$, $v_\phi^\theta$). Visual inspection shows the rotation angle $\theta$ to be between $10\deg$ and $20\deg$ depending on the band and radius considered. To simplify the method we use the same angle for all bins and we choose a value of $15\deg$ for reasons specified in step \ref{i:marginalize}.

\item \label{i:density}We estimate the probability density in this velocity space using the Epanechnikov adaptive kernel density estimator method \citep{Silverman86} with an adaptability exponent of 0.1. 
(Fig.~\ref{findvolrsim} second column).  %!! talk a bout biases of that. using always kernel?

\item \label{i:marginalize}We integrate over $v_R^\theta$, only inside the range $v_R^\theta$=[-130,-10]~$\kms$ (within the black vertical lines in the second column) to avoid contamination from other groups or regions of the velocity plane. The distribution along $v_\phi^\theta$ is shown in the third column. We clearly see the presence of the two peaks separated by a valley. The Hercules peak is indicated with a black vertical line. We see how for small $R$, Hercules is stronger than the MAIN mode, while as we move outwards in the disc it becomes weaker. Of all rotation angles $\theta=10, 15, 20, 25\deg$,  the angle of $\theta=15\deg$ gives the maximum height of Hercules (black vertical line) for most of the bands and radial bins. This means that for this angle the structures are better aligned with the horizontal axis. In step \ref{i:invrot} we estimate the error on the final location of the saddle point derived by assuming this value. Then we locate the position of the minimum $\volrrot$ (red line). 

\item \label{i:maxmin}We estimate the error in $\volrrot$ by generating 500 bootstrap samples, repeating steps \ref{i:rotate} to \ref{i:marginalize}, and computing 
the standard deviation of the obtained set of $\volrrot$, which is typically very small ($\sim 2\kms$, red error bars in Fig.~\ref{findvolrsim} third column).
Fig.~\ref{vRsim} (second panel) shows $\volrrot$ as a function of $R$ for the 5 radial bins (blue diamonds). These velocities decrease with $R$. Additionally, we show $\volrrot$ for the other bands at $20\deg$, $60\deg$ and $80\deg$ in different colours which depict the same behaviour.

  \begin{figure}
   \centering
 \includegraphics[width=0.4\textwidth]{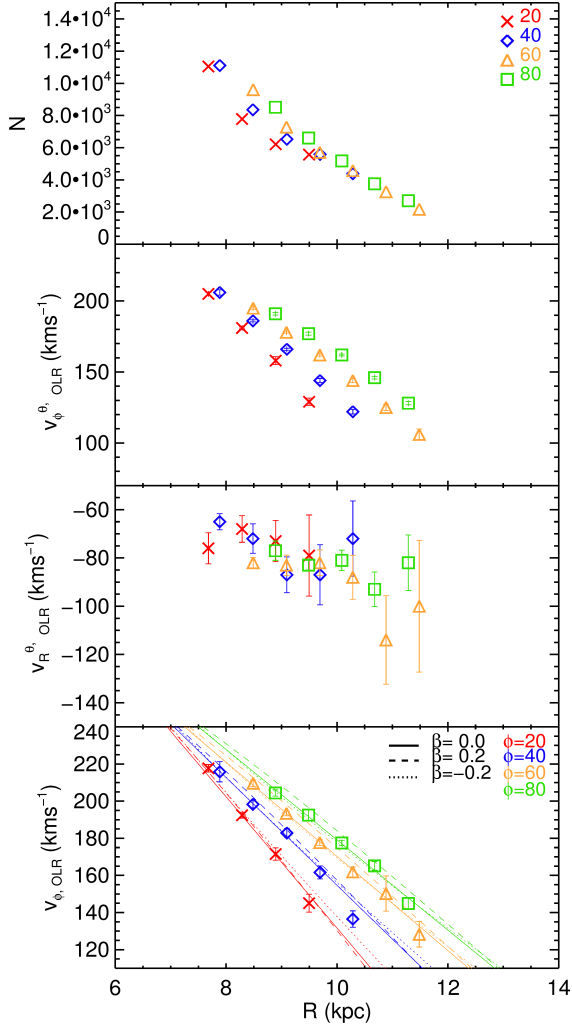}
  \caption{Several measurements for the bands at different bar angles of the simulation as a function of radius $R$. Top panel: number of stars per bin. Second panel: measured $\volrrot$. Third panel: measured $\uolrrot$. Bottom panel: final determination of $\volr$ and expected theoretical curves.}
         \label{vRsim}
   \end{figure}

\item \label{i:m1}To get the $v_R^\theta$ of the saddle point, we derive the distribution along $v_R^\theta$ (fourth column) inside the green rectangles of Fig.~\ref{findvolrsim}. These are centred in the valley ($\volrrot$) with a width of $20\kms$. We then find the maximum of the curve $\uolrrot$ (green line). 
We also estimate the error in $\uolrrot$ through the bootstrapping technique and it is typically of $5-10\kms$ (green error bars). The position of $\uolrrot$ as a function of $R$ is shown in the third panel from the top of Fig.~\ref{vRsim} (blue diamonds). In general, this velocity becomes more negative with radius. For larger radii the distribution in $v_R^\theta$ is noisier and has several maxima (bottom panels of Fig.~\ref{findvolrsim}). This is because the number of particles for large $R$ decreases and the valley is wider and contains less particles (only 24 particles were inside the green rectangle of the last radial bin). In this case our determination of $\uolrrot$ may not be accurate. For instance, the last bin of the band at $40\deg$ does not follow the overall trend.
%\item \label{i:m2} An alternative method to determine $\uolrrot$ is to use the value of the maximum density along the Hercules structure itself. This seems to be approximately aligned to the saddle point. To determine this maxima, we simply compute the  $\vr$ of the particles located in the structure, selecting them inside a rectangle centred in the determined position $\vphi$ of Hercules in step \ref{i:maxmin}
%(orange rectangle in second column) with a width in $\vphi^\theta$ of $20\kms$ in the direction of decreasing $\vphi^\theta$ and $10\kms$ in the direction of increasing $\vphi^\theta$.
%These alternative determinations are shown as orange vertical dashed lines in the fourth column in Figure \ref{findvolrsim} and we see that both determminations are similar. Also third panel of Figure \ref{vRsim} shows the obtained values as a function of $R$ for the different bands, which shows a more thigh relation for large radii, where the differences between the two determinations are larger. We see how this second determination gives smaller values for small $R$ but larger values for larger $R$. We see that for $68\%$ of the bins in all bands, the difference between the two determinations of $\uolrrot$ is smaller than $5.8\kms$. And nevertheless, a difference of $5.8\kms$ in $\uolrrot$ translates into a difference in our final determination of $\volr$ (for which a rotation of $-\theta$, see step \ref{i:invrot}) of only $1.98\kms$.

\item \label{i:invrot}We convert ($\uolrrot$, $\volrrot$) back to ($\uolr$, $\volr$) by rotating an angle $-\theta$. The position of the saddle point is indicated with red bars in the first and second columns of Fig.~\ref{findvolrsim}. The value $\volr$ is the observable needed in our modelling. To obtain the errors in $\uolr$ and $\volr$ we must consider two contributions. First, the statistical errors which, as explained above, we get from the bootstrapping method ($e_{bst}$). Second, the error made by using a fixed value for the rotation angle $\theta$ ($e_{\theta}$). To estimate the latter we repeat steps \ref{i:rotate} to \ref{i:invrot} using the two extreme angles of $\theta=15\pm5\deg$, we compute the maximum difference between the new determinations of $\uolr$ and $\volr$ and the ones for $\theta=15\deg$ and assign this difference to the error, which turns out to be $\lesssim 5\kms$. Finally, we add both errors $e_{bst}$ and $e_{\theta}$ in quadrature.
\end{enumerate}

The measured velocities $\volr$ are shown Fig.~\ref{vRsim} (bottom) for the four bands. For all bands, the velocity decreases with $R$. Overlaid on the points are the theoretical curves from Eq.~\ref{e:fitR} for the input parameters of the simulation ($\Ob=47.5\kmskpc$, $\vo=220\kms$ and $\Ro=8.5\kpc$), along with the four bar orientations of the bands. As the slope of the rotation curve changes with radius for the A91 model (Sect.~\ref{test0}), we plot for each band three curves corresponding to $\beta=0.2, 0, -0.2$. We see that our measured $\volr$ are consistent with the predictions of Eq.~\ref{e:fitR}, given the errors. We see more discrepancies at large radii where it is more difficult to detect reliably the position of the saddle point for the reasons mentioned above. Note that the discrepancy between the estimated value in
the last radial bin for the band at $40\deg$ and that expected is due to the poor determination of $\uolrrot$ (step \ref{i:m1}) and we shall reject this data point in our analysis of Sect.~\ref{results0}. 

We have also validated the analytical model with other simulations with different values of the pattern speed, which moves the resonances to other positions of the disc, and the bar's force and obtained similarly satisfactory results. We will now recover the input model parameter from the simulated data.

%Comment also on analysis with: few number of particles, with wavelet transform  ?

%__________________________________________________________________
\subsection{Maximisation and parameter space sampling}\label{maximisation}
%-------------------------------------------------------------

We compare the determined values of the position of the saddle point ${{\volr}_{i}}$ with the estimates obtained from
 Eq.~\ref{e:fitR} through the chi-square statistic:
\beq
\chi^2=\sum_{i}\Bigg(\frac{{{\volr}_{i}}-{{\volr}_{i}}^{model}}{{{\sigma}_{i}}}\Bigg)^2
\eeq
where the subscript $i$ stands for each data point and ${\sigma}_{i}$ is the error of each point. We assume that the noise associated with the data points can be represented as a Gaussian process and, therefore, we can approximate the likelihood function by
$prob \propto \exp\left(-\frac{\chi^2}{2}\right)$. In the maximisation of the probability, we consider the pattern speed $\Omega_b$ in units of $\Oo$, i.e. in practice we fit $\Omega_b/\Oo$. We take a range of $[0, 3.4]\Oo$, which corresponds to $[0, 100] \kmskpc$, in steps of $0.0025$ or $\sim0.07\kmskpc$. This range is large enough not to
influence the posterior probability density function (pdf). For the slope of the rotation curve $\beta$ we use the range $[-0.2, 0.2]$ in steps of $0.01$. These limits are the ones considered in D00 for which the fit given in Eq.~\ref{e:fit} is valid. The bar's angle $\pb$ is explored in the range of $[0, 80] \deg$ in steps of $0.5\deg$. Outside this range, the model of Eq.~\ref{e:fit2} is not valid as the Hercules structure does not exist or there is just a counterpart at $v_R>0$ in the velocity plane (e.g. Fig.~2 in D00). Note that this range is actually larger than the limits considered in D00 to obtain the fit of Eq.~\ref{e:fit} (he used up to $50\deg$). However, as we showed in Sect.~\ref{finding0} (Fig.~\ref{vRsim}) the extrapolation to larger angles is valid. 

%__________________________________________________________________
\subsection{Recovering the parameters of the model}\label{results0}
%-------------------------------------------------------------

\begin{table*}
\caption{Results of the fits for the toy model. The input pattern speed is $\Ob=1.836\Oo=47.5\kmskpc$ for all cases. }             
\label{t:fitssim}      
\centering          
%\begin{tabular}{l l l l l l l l l ll }     % 11 columns 
    \tabcolsep 3.pt
\begin{tabular}{l lccccccccc }     % 11 columns 
\hline\hline       
\multicolumn{2}{l}{Input}&\multicolumn{1}{c}{$\pb$(MAX)}&\multicolumn{1}{c}{E($\pb$)}&\multicolumn{1}{c}{$\sigma_{\pb}$}&\multicolumn{1}{c}{$\Ob/\Oo$(MAX)}&\multicolumn{1}{c}{E($\Ob/\Oo$)}&\multicolumn{1}{c}{$\sigma_{\Ob/\Oo}$}&\multicolumn{1}{c}{$\rho_{\pb\Ob}$}&\multicolumn{1}{c}{E($\Ob/\Oo|\pb=input$)}&\multicolumn{1}{c}{E($\Ob|\pb=input$)}\\ 
\multicolumn{2}{l}{}&($\deg$)&($\deg$)&($\deg$)&&&&&&($\kmskpc$)\\ 
%&&&$(\kpc)$&$(\kms)$&$(\kms)$&&&$(\kmskpc)$&$(\deg)$\\ 
\hline      
 $20\deg$&&10.&    11.&     6.&     1.80&     1.80&     0.02&     0.92&     1.83$\pm$     0.01&    47.4$\pm$     0.2\\
 $40\deg$&&29.&    32.&    12.&     1.79&     1.80&     0.06&     0.98&     1.84$\pm$     0.01&    47.7$\pm$     0.3\\
 $60\deg$&&80.&    55.&    13.&     1.90&     1.81&     0.06&     0.97&     1.83$\pm$     0.02&    47.4$\pm$     0.4\\
 $80\deg$&&80.&    65.&    10.&     1.84&     1.78&     0.05&     0.90&     1.85$\pm$     0.02&    47.9$\pm$     0.5\\
\hline                  
\end{tabular}
\end{table*}

\begin{figure}
   \centering
 \includegraphics[width=0.4\textwidth]{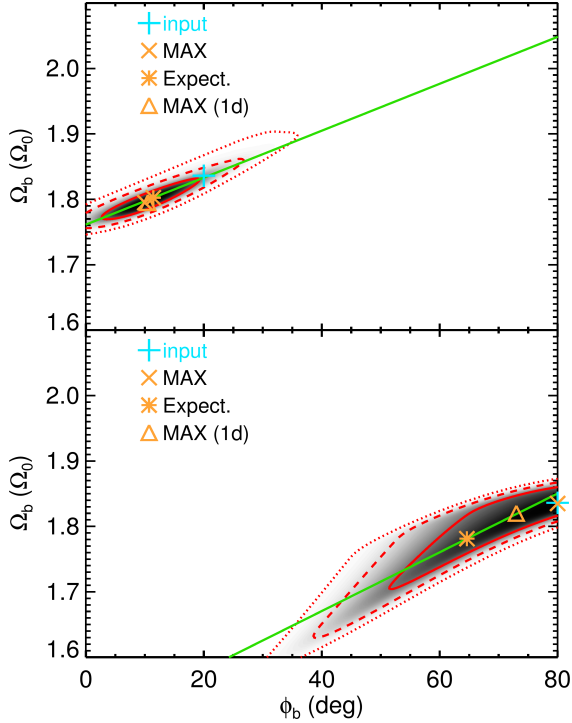}

  \caption{Two-dimensional marginalised likelihood (over the parameter $\beta$) of the model parameters for the simulated bands at $40\deg$ (top) and $60\deg$ (bottom). Solid, dashed and dotted red lines show the $1\sigma$, $2\sigma$ and $3\sigma$ confidence regions, respectively. The input parameters of the simulation are shown with a light blue cross. The maximum and the expected value of the two-dimensional pdf are indicated with a orange cross and asterisk, respectively. The orange triangle is the maximum of the one-dimensional marginalised pdf's.}
         \label{fitmodel2d}
   \end{figure}
%
%  \begin{figure}
%   \centering
% \includegraphics[width=0.24\textwidth]{fig8a}
% \includegraphics[width=0.24\textwidth]{fig8b}
%
% \includegraphics[width=0.24\textwidth]{fig8c}
% \includegraphics[width=0.24\textwidth]{fig8d}
%
% \includegraphics[width=0.24\textwidth]{fig8e}
% \includegraphics[width=0.24\textwidth]{fig8f}
% \caption{Marginalized likelihoods of $\Ob$ (top), $\pb$ (middle),  and $\beta$ (bottom) for the simulated bands at $40\deg$ (left) and at $60\deg$ (right). Dotted black and red error bars show the best estimate and error taken as the maximum likelihood value and the $1\sigma$ confidence interval, and the mean and standard deviation of the pdf, respectively. The blue dashed lines correspond to the likelihood for the prior on $\pb$ (see text).}
%         \label{fitmodel1d}
%   \end{figure}

Figure \ref{fitmodel2d} shows the pdf for our toy model in the $\pb$-$\Ob$ plane for the bands at $40$ and $60\deg$. This pdf has been marginalised over $\beta$, since we
do not expect to constrain the slope of the rotation curve $\beta$ to a single value as it varies in our simulations 
over the distance range considered. Indeed, we find a very flat pdf in the direction of $\beta$. 
In Table~\ref{t:fitssim} we give the maximum of the probability, the mean or expectation of each parameter E$(\pb,\Ob)$, together with the standard deviation of the pdf $\sigma_{\pb}$ and $\sigma_{\Ob/\Oo}$, and the correlation coefficient $\rho_{\pb\Ob}$ for each band. In Fig.~\ref{fitmodel2d} the $1\sigma$, $2\sigma$ and $3\sigma$ confidence regions are delimited by the dotted, dashed and solid red lines, respectively. The maximum of the probability is indicated with a orange cross whereas the mean E$(\pb,\Ob)$ is shown with a orange asterisk. There is a high correlation between orientation and pattern speed (also noticed in D00) with a correlation coefficient around $\rho_{\pb\Ob}\sim0.98$, with higher values of $\Ob$ preferred for larger bar angles.

For the band at $40\deg$ the maximum and the mean of the pdf are similar. They are also close to the input value of the model (light blue cross), which is in the limit between the $1\sigma$ and $2\sigma$ confidence regions. However, if we remove the (problematic) last bin in $R$ for this band, the input value lies well inside the $1\sigma$ region. For the band at $60\deg$, the mean and the maxima of the pdf differ significantly. This is because the probability distribution is flatter and more asymmetric, and in this case the mean can be considered a better estimate and more representative as it takes into account the skewness of the pdf. The input value is close to the mean value and falls well inside the $1\sigma$ region. From the values in Table~\ref{t:fitssim} we see that the recovered values can present an offset with respect to the input parameters of around $\sim10\deg$ in the orientation but only of $0.04\Oo$ or $\sim1\kmskpc$ for the pattern speed. Nevertheless, input and recovered values are 
consistent given the standard deviations. We obtain similar results for the other bands.

%The solid black lines in Fig.~\ref{fitmodel1d} show the pdf obtained for the individual parameters marginalizing over the rest of them. The maximum likelihood values together with the $1\sigma$ limits are indicated with a black symbol with an asymmetric error bar. In red we plot the mean of the pdf and its standard deviation as error bars. The dashed vertical lines are the input parameters of the simulation. The top panels show that we obtain a rather defined peak for the pdf of $\Ob$ (note that in the plot we have reduced the initial maximisation range for this parameter [$0$, $3.4$]$\Oo$). 
%For the bar's orientation (second row), we obtain broader pdfs.
% although the input values are still inside the $1\sigma$ limits for the band at $40\deg$. For the other band at $60\deg$ the orientation of the bar is not so well constrained probably due to the fact that, fixed the pattern speed, the curves for large bar angles are more close to each other (see Figure \ref{vRsim}, bottom panel). 
%Finally, we obtain a flat pdf for $\beta$ (third row) as expected.

The two-dimensional pdf contours are approximately elliptical and can be locally well approximated by a multivariate Gaussian centred on the expected values and with a covariance matrix given by the values of Table~\ref{t:fitssim}. This approximation allows us to establish a tighter joint constraint on the set of parameters (orientation and pattern speed). Furthermore, if we have independent constraints on the bar's orientation $\pb={\pb}_1$, our conditional best estimate for $\Ob$ would be:
\beq\label{e:linear}
E(\Ob/\Oo|\pb = {\pb}_1) =E(\Ob/\Oo) + \frac{
\rho_{\pb\Ob}\sigma_{\Ob/\Oo}
}{\sigma_{\pb}}
\left({\pb}_1-E(\pb)\right)
\eeq
with a variance:
\beq\label{e:variance}
Var(\Ob/\Oo|\pb = {\pb}_1) ={\sigma^2_{\Ob/\Oo}}\left(1- {\rho^2_{\pb\Ob}}\right)
\eeq
This linear relation is shown as a green line in Fig.~\ref{fitmodel2d}. For example, we might put a prior on $\pb$ to be the exact input value, i.e. $40$ or $60\deg$. The resulting conditional expected values are indicated in the last two columns of Table~\ref{t:fitssim} (in units of $\Oo$ and in $\kmskpc$) and we see that we recover with high accuracy ($1\%$) the input pattern speed. This can also be seen in Fig.~\ref{fitmodel2d} where the light blue cross almost lies on top of the green line.

If we marginalise the pdf's of Fig~\ref{fitmodel2d}, we obtain the best estimates for each individual parameter independently on the rest of parameters and their corresponding confidence intervals. The maxima of the individual marginalisations are shown with a orange triangle in Fig.~\ref{fitmodel2d}. Whereas for the band at $40\deg$, this yields a peak that is close to the maximum in the two-dimensional pdf, for $60\deg$ the new peak is completely off. This is because the global pdf is highly degenerate and asymmetric, and the one-dimensional pdf's do not capture the main correlation between the parameters, giving unsatisfactory results. Therefore our best results are given when the two parameters are simultaneously estimated. 

%__________________________________________________________________
\section{Hercules in the RAVE data}\label{application}

We use now the RAVE survey data to measure the position of the saddle point $\volr$ as a function of R. 
Our aim is to establish whether the observed trend is consistent with the analytic model developed in Sec.~\ref{model} and to constrain the bar properties through best fits to the observations.

\subsection{The RAVE data}\label{data}

We use the RAVE DR4 \citep{Kordopatis13} and the distance determination method by \citet{Zwitter10}, which leads to a new data set with 315572 stars\footnote{In A12 we used DR3 and distances by \citet{Burnett11} with 202843 stars with 6D phase-space information.} Alternatively, we may also use the distances from the method by \citet{Burnett11}, presented in \citet{Binney13} which are in fact the recommended ones for DR4. We shall see later, however, that we obtain very similar results in both cases. The stellar atmospheric parameters of the DR4 are computed using a new pipeline, based on the algorithms of MATISSE and DEGAS, and presented in \cite{Kordopatis11}. Compared to DR3, DR4 is 5 times larger and the spectral degeneracies and the 2MASS photometric information are better taken into consideration, improving the parameter determination (and hence the distance estimation) with respect to previous data releases. 
We use proper motions from different catalogues, mainly PPMX
(\citealt{Roser08}) and UCAC2 (\citealt{Zacharias04}), choosing from each catalogue the values with the smallest errors.

\begin{figure}
   \centering
   \includegraphics[width=0.35\textwidth]{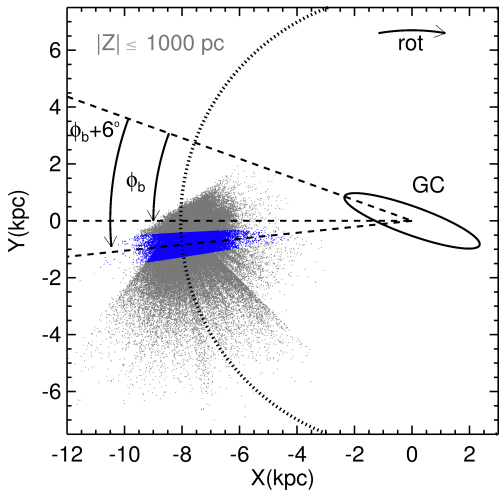}
  \caption{Positions of the RAVE DR4 stars selected with $|Z|\leq1\kpc$ (grey dots) together with the stars selected in the band at $\pb+6\deg$ with respect to the bar (blue dots). The Sun is at X=-8.05 and Y=0. A schematic bar with an (arbitrary) orientation of $\pb=20\deg$ is also shown.}
         \label{xyRAVE}
   \end{figure}

 Following \citet{Honma12} we use a position of the Sun of $(X,Y)=(-8.05,0)\kpc$ and a circular velocity at the Sun of $\vo=238\kms$ 
to compute the positions and cylindrical velocities $\vr$ and $\vphi$ of the stars. 
 We adopt the velocities of the Sun with respect to the Local Standard of Rest of $(\Us, \Vs, \Ws)=(10,12,7)\kms$ from \citet{Schonrich10}.%\citet{McMillan10}
We examine later on the implications of these adopted values on our results. Figure \ref{xyRAVE} (grey dots) shows the positions of these selected RAVE stars.
The value of $(\vo+\Vs)/\Ro$ by \citet{Honma12} is $31.09\pm0.78  \kmskpc$ which is compatible with the value from the reflex of the motion of the Sgr A* $30.2 \pm 0.2 \kmskpc$ \citep{Reid04}.

  \begin{figure}
   \centering
 \includegraphics[width=0.45\textwidth]{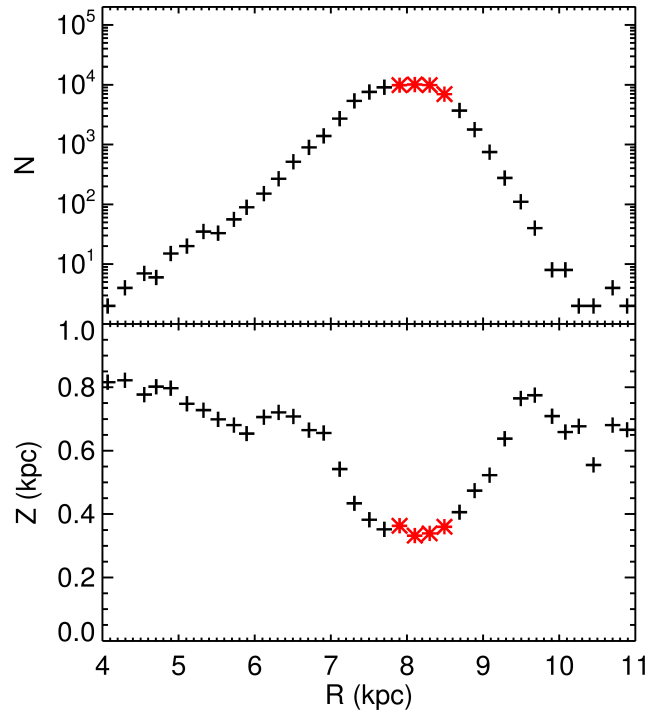}
  \caption{Number of stars per bin in $R$ (top) and median $Z$ coordinate (bottom) as a function of $R$ for the band selected in the RAVE DR4. The red asterisks are the bins used in our analysis.}
         \label{regions}
   \end{figure}

First, we select 274103 stars in the plane with $|Z|\leq1\kpc$, as for this range of heights we expect to be able to detect the effects of the bar \citep{Monari13}.
From the stars within $1\kpc$ from the plane, we select a band of stars at $6\deg$ with respect to the line Sun-GC with a width of $\Delta\phi=6\deg$ (blue dots in Fig.~\ref{xyRAVE}). There are in total 71605 stars in this band, of which 94\% are giants. The median relative error in distance for this band is $27\%$, the median error in transverse velocity is $20\kms$, whereas radial velocity errors are smaller than $1.5\kms$ for $90\%$ of the stars. If $\pb$ is the orientation of the bar with respect to the Solar neighbourhood, the orientation of this band with respect to the bar is $\pb+6\deg$.
%The bands are also selected to have $|Z|\le 1 \kpc$ but we will discuss the results also for other heights. %check that we do
%We also consider separately giant and dwarf stars, making it easier to interpret the effects of the possible different biases in the distance determination.; not anymore 

 We choose this band because it covers a large range of $R$ while keeping the errors in distance and kinematics small.  Ideally, one could use data on different bands, i.e. not restricted to a given $\phi$. However,  
  given the quantity and quality of the current data, this does not improve the fit on the parameters: bands at other angles have less stars and larger kinematic errors or cover a smaller range of radii. 
  First, as one can see in Fig.~\ref{xyRAVE}, the RAVE data extends far beyond the blue band selected. However, at these locations the Hercules stream is hardly recognisable. The stars are at least $1.5-2.\kpc$ far from the Sun and their distance and kinematic errors are large. One could also take a band for the $\phi$ of the Solar Neighbourhood. However, 
  the data for this band does not cover a large range of $R$ and is so close in angle (only $6\deg$) to the band selected that we do not improve the fit by combining the information of the two bands.

As in Sec.~\ref{test}, we divide the band in bins of $R$ but 
we now take bins every $0.2\kpc$ with a width of $\Delta R=0.2\kpc$. In Fig.~\ref{regions} (top) we show the number of stars per bin.
The range of radii that we probe for this band is $[7.8,8.6]\kpc$ (red asterisks in the figure). 
Outside this range we fail to detect the Hercules structure. This may be due to observational errors, to the fact than Hercules is masked by the other groups and due to the number of stars which decreases substantially. On the other hand, the average height of the stars (bottom panel of Fig.~\ref{regions}) increases significantly outside the mentioned range due to the RAVE fields selection. For large heights above the plane the kinematic structures may be also diluted and, additionally, the behaviour of the orbital frequencies can be different from those in-plane which may invalidate Eq.~(\ref{e:fitR}).

%__________________________________________________________________
\subsection{Measuring $\volr$ in the RAVE data}\label{finding}

  \begin{figure*}
   \centering
   \includegraphics[height=0.45\textheight]{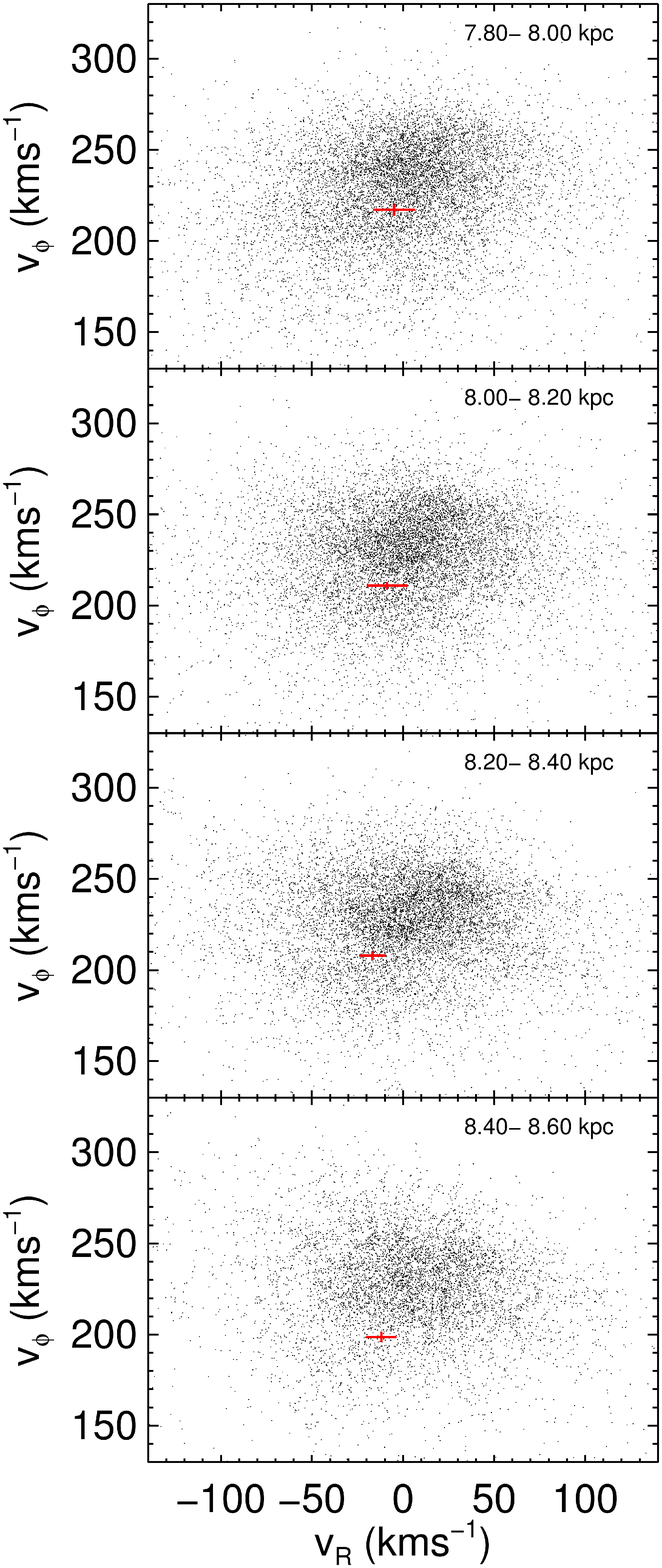}
   \includegraphics[height=0.45\textheight]{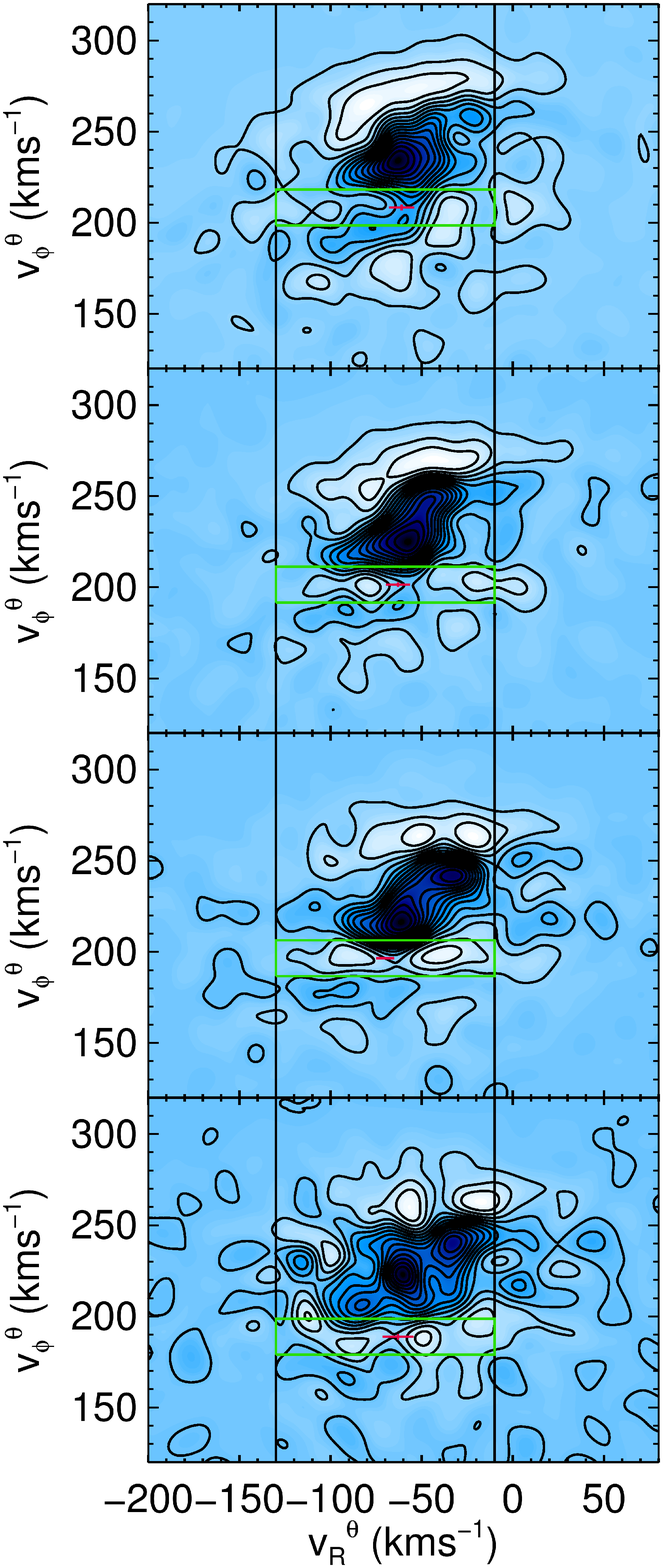}
   \includegraphics[height=0.45\textheight]{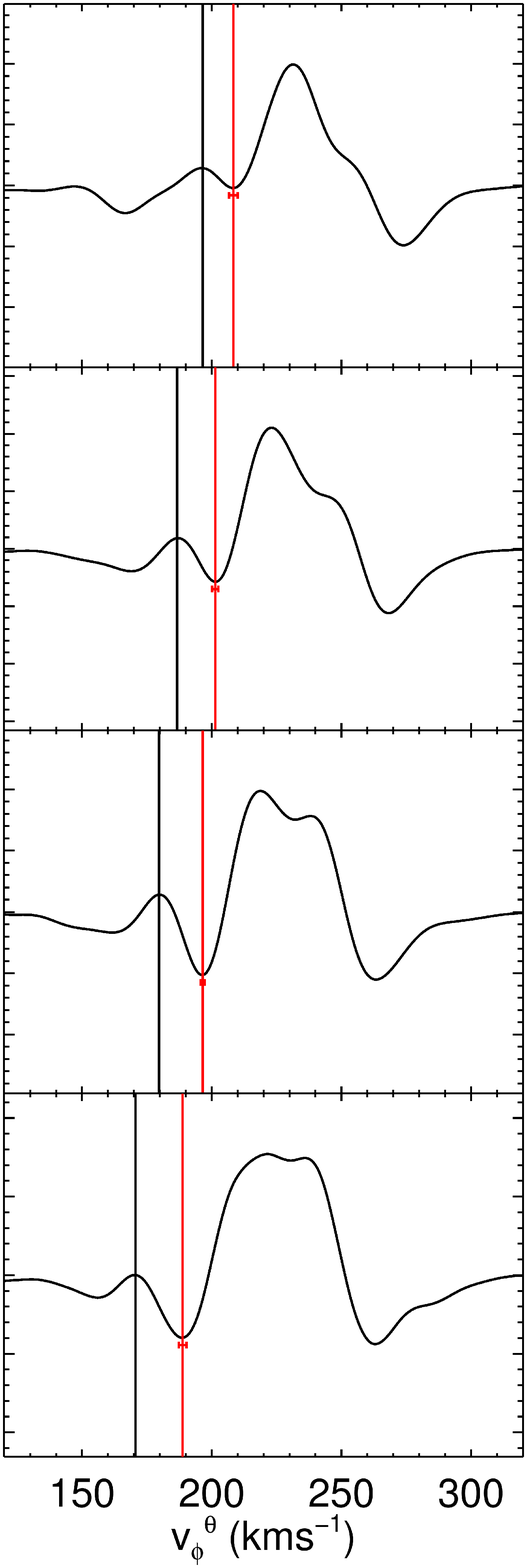}
   \includegraphics[height=0.45\textheight]{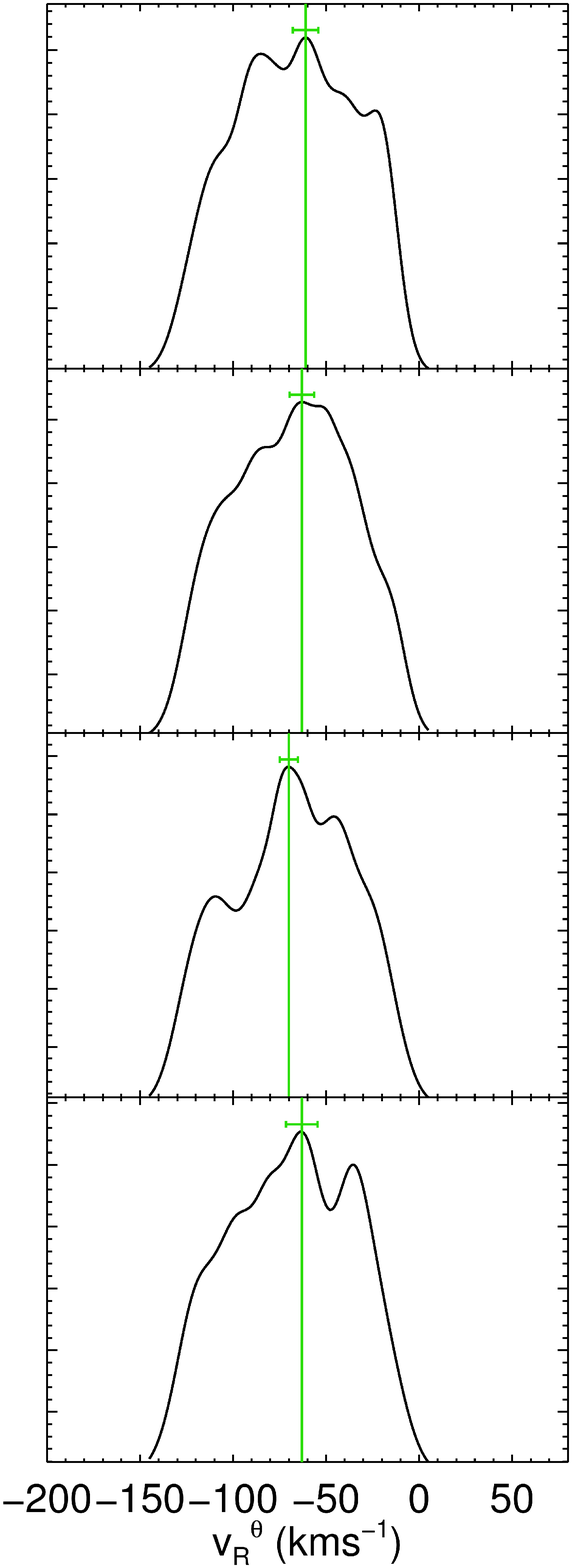}
  \caption{Same as Fig.~\ref{findvolrsim} but for the band at $\pb+6\deg$ for the RAVE data. The velocity distribution of the second column is obtained through the wavelet transform (see text).}
        \label{findvolr}
  \end{figure*}

We now follow the steps outlined in Sect.~\ref{finding0} using the RAVE data. The process is shown in Fig.~\ref{findvolr}. % and the presence of other groups that mask the Hercules stream. 
 The first column shows that the Hercules stream is not as clear as in the simulations depicted in Fig.~\ref{findvolrsim}. This may be due to several reasons. First, the presence of observational errors dilutes velocity structures. Also the existence of other kinematic groups in the data may mask the presence of Hercules. Second, the bins in the RAVE data are located at relatively large heights from the plane (Fig. \ref{regions}) which can also wash out the Hercules signal. Third, the test particle simulation of Sect.~\ref{test} shows a particularly conspicuous OLR mode but \citet{Monari13} have shown recently that for longer integration times (i.e. older bars) the distinction between the MAIN and the OLR modes is less clear and more similar to observations. The strength of the bar can also influence in the conspicuousness of Hercules.

\begin{figure}
   \centering
\includegraphics[width=0.4\textwidth]{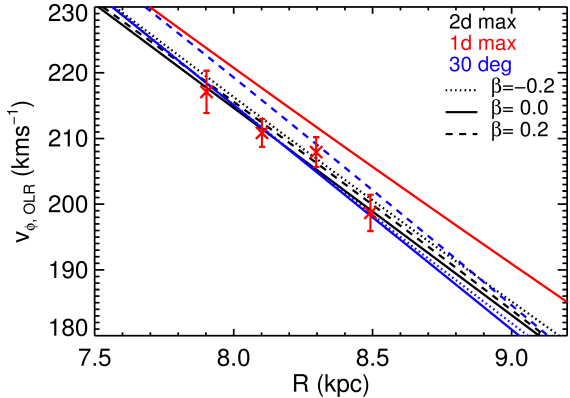}
 \caption{Position of the $\volr$ as a function of $R$ for the band at $\pb+6\deg$ for the RAVE data. Several fits from Table \ref{t:fits} (see text) are overplotted.}
         \label{vRdata}
   \end{figure}

Due to the above limitations we introduce a change in our method with respect to Sect.~\ref{test}. This is because the adaptive kernel density estimator produces a weak signature of the Hercules peak, sometimes seen only as an inflection point. We therefore prefer to use the wavelet transform (WT) instead. This is especially suitable to enhance overdensities and underdensities (Fig.~\ref{findvolr} second column), and has been applied extensively for the detection of kinematic groups (\citealt{Skuljan99,Antoja08}, A12)\footnote{For our simulations of Sect.~\ref{test}, we also tried the WT but concluded that the kernel density estimator performed better. The reason is that the WT overestimates the position of the gap $\volrrot$ for the cases where the Hercules structure is remarkably separated from the MAIN mode or, in other words, where the gap is wider than $60\kms$ (for bins at the outermost radii). As this is not the case of any bin of the RAVE data, we are not affected by this WT bias here.}. We use here 
a range of scales between $22-45\kms$ (see A12). The WT detects also other peaks such as Hyades or Sirius apart from the Hercules group in the distribution of $\vphi^\theta$ (third column in Fig.~\ref{findvolr}). 
%as the distributions in the fourth column do not allow us to detect a defined maximum. The middle panel of Figure \ref{vRdata} shows the determination of $\uolrrot$ for the different bins. A slightly decreasing trend is observed as in the simulations. 
%To estimate the error in the determination of $\volr$ and $\uolr$ we now have to consider also the errors in the observables (distances, proper motions and radial velocities). To do this, instead of using the bootstrapping technique, we
%generate 300 versions of the whole RAVE sample after a Gaussian error convolution directly on the observables. Then, we select the stars in the $\Delta \phi$ band and repeat the process of finding $\volr$ and $\uolr$. Finally we compute the dispersion for the set of 300 samples 
%%Although ideally one would like to deconvolve the data with the observational errors, w
% and use it as an approximated measure of the effects of observational errors in our determination of $\volr$.% This process probably overestimates the errors 
%%mainly because i) we are doing a second convolution with errors, ii) the number of stars per bin is reduced between $5$ and $10\%$,% only checked for this band
%%and iii) in some cases, the error convolution dilutes the structures and the saddle point is not found or it is confused with other peaks due to noise. 
Figure \ref{vRdata} shows the $\volr$ for the different bins which decreases with $R$ as expected if Hercules is caused by the bar's OLR. The data points are also tabulated in Appendix \ref{tables}.

%__________________________________________________________________
\section{Results: application of the analytic model to the RAVE data}\label{results}

We proceed to obtain the most likely bar properties consistent with the RAVE data. We use the maximisation parameter ranges as explained in Sect.~\ref{maximisation}. 
Note that we had to assume values for $\vo$, $\Vs$, $\Us$ and $\Ro$ to compute the individual $\vphi$, $\vr$ and $R$ from the observables. In Sect.~\ref{results1} we keep these parameters fixed, while in Sect.~\ref{results2} we consider also changes in these parameters. In Sect.~\ref{results12} we discuss the effect of the observational errors and possible biases in distance.

%and we are not aware of independent determinations for this parameter being $>100.\kmskpc$. 

%%%%%%%%%%%%%%%%%%%%%%%%%%%%%%%%%%%%%%%%%%%%%%%%%%%
\subsection{Results for fixed Solar parameters}\label{results1}

\begin{figure*}
   \centering
 \includegraphics[width=0.325\textwidth]{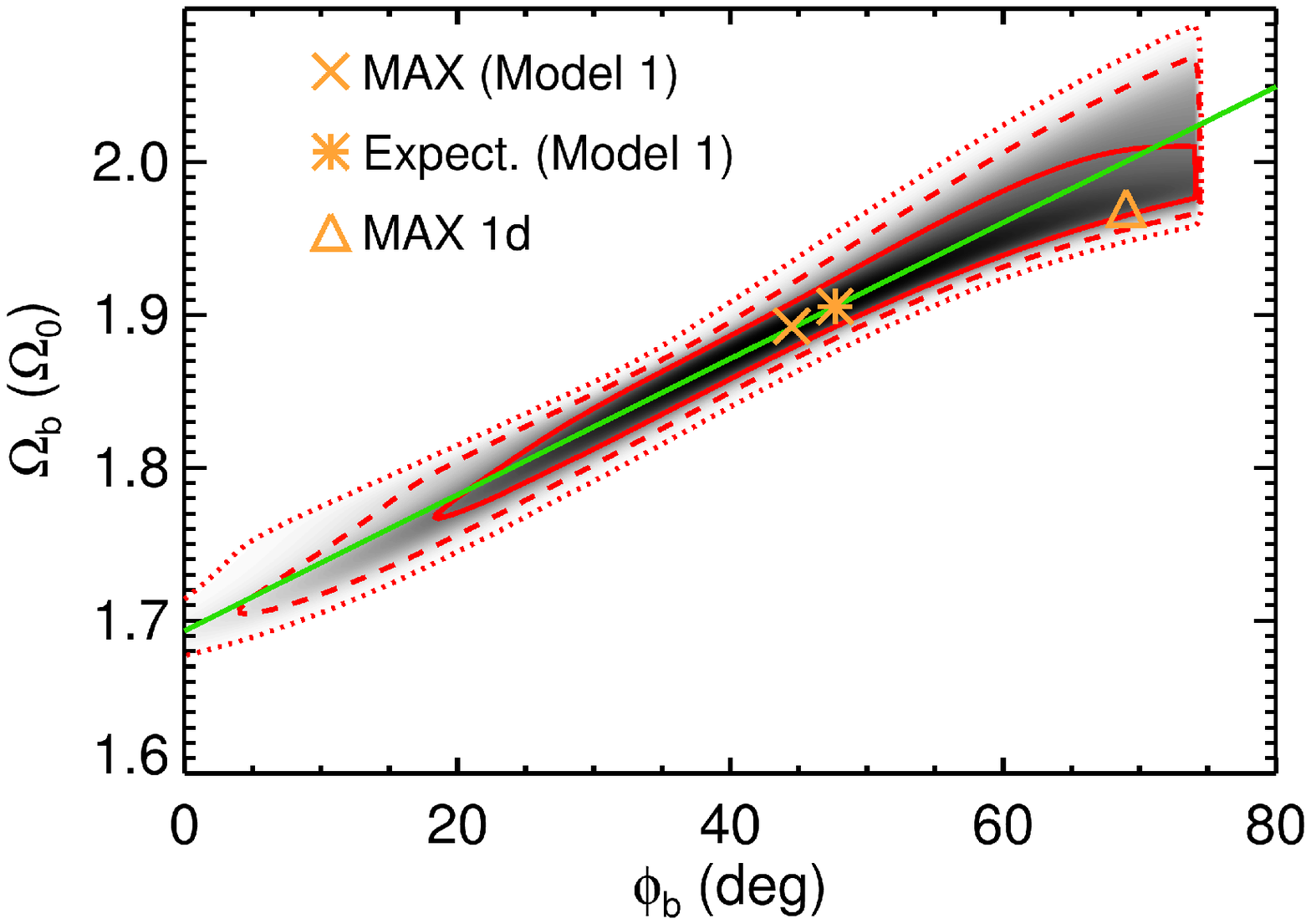}
 \includegraphics[width=0.325\textwidth]{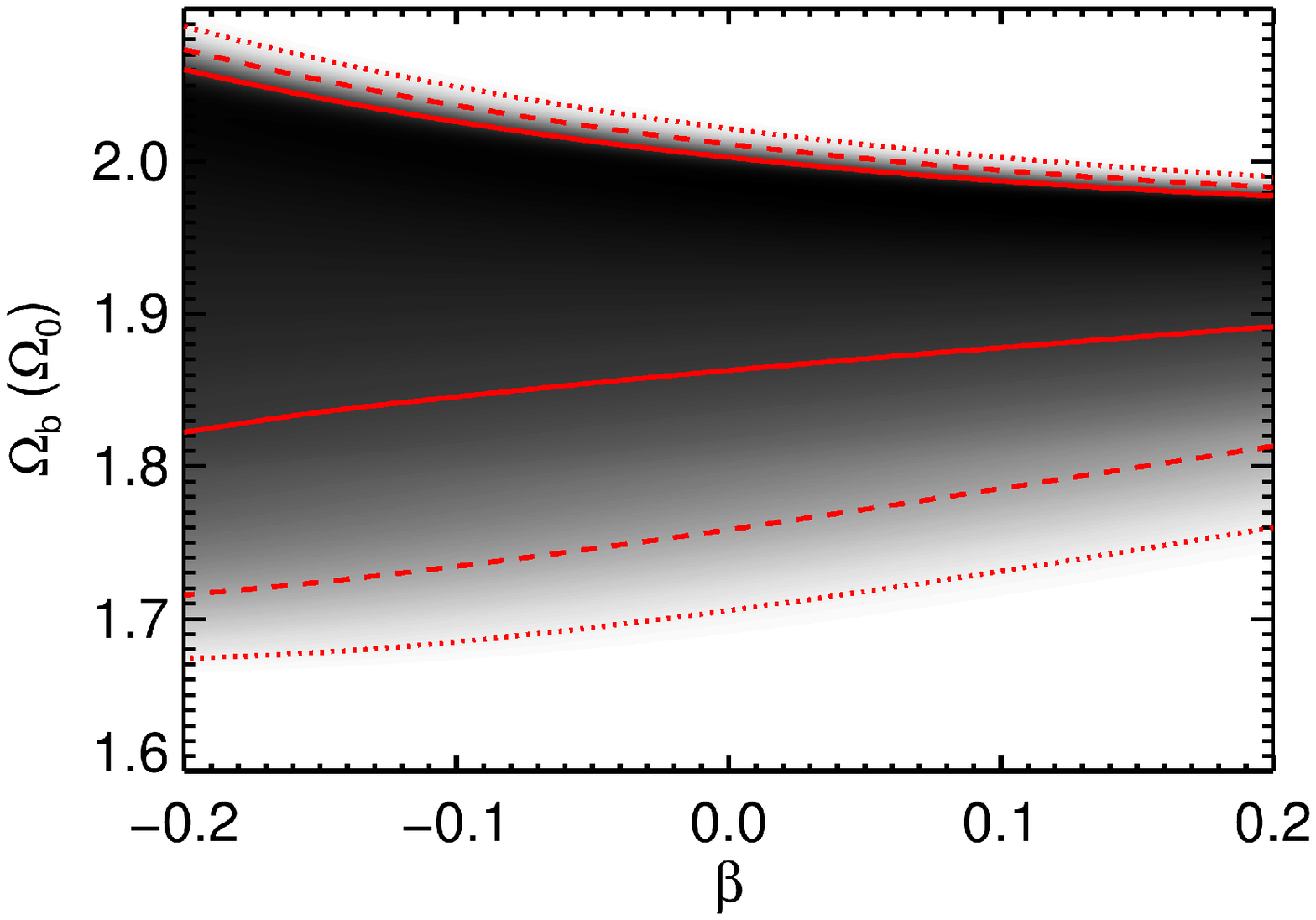}
 \includegraphics[width=0.325\textwidth]{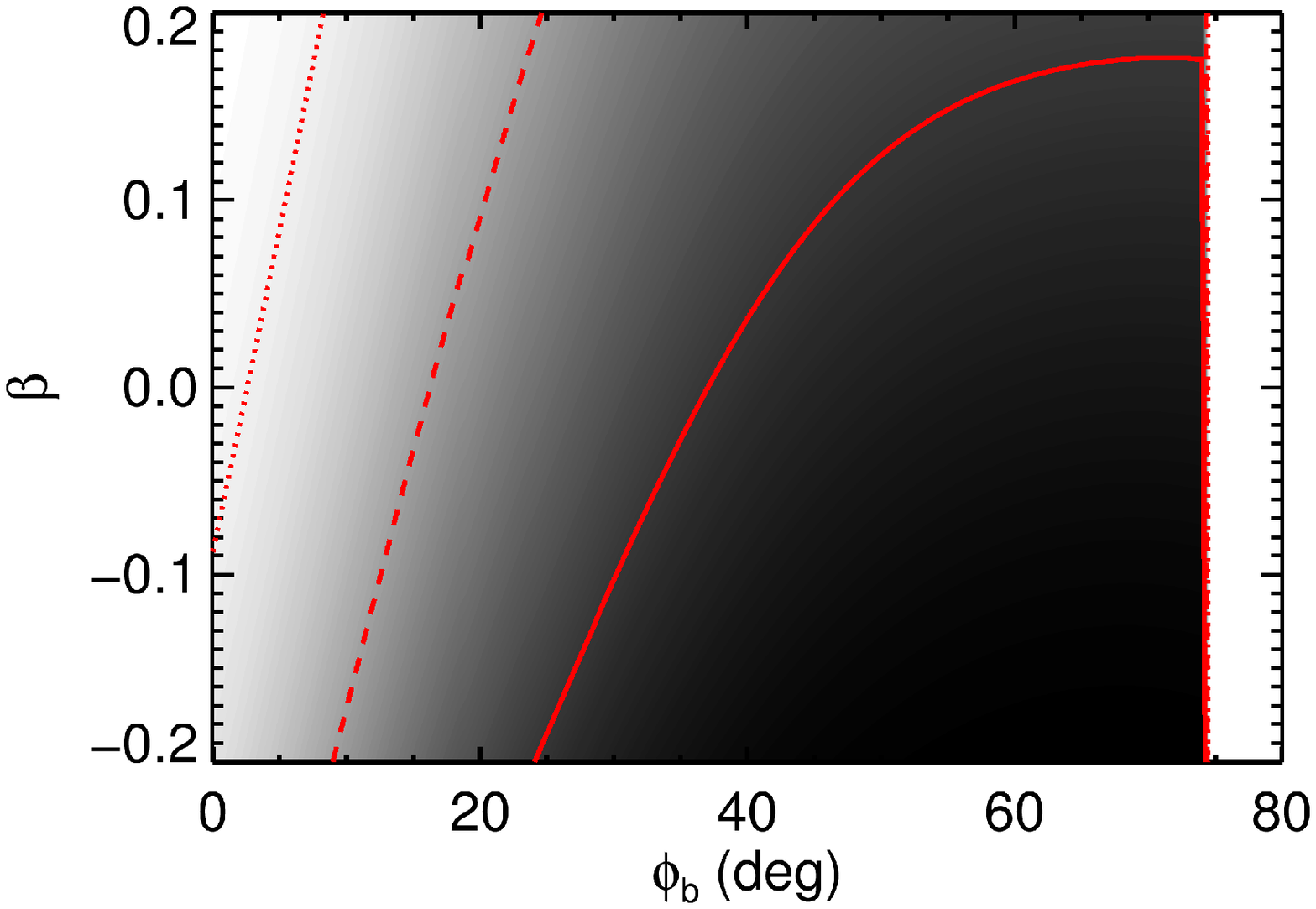}
  \caption{Two-dimensional marginalised likelihoods (over the 3rd parameter) for the model parameters for the RAVE data. Dotted, dashed and solid red lines show the $1\sigma$, $2\sigma$ and $3\sigma$ confidence regions, respectively. In the left panel the maximum and the expected value of the two-dimensional pdf are indicated with a orange cross and asterisk, respectively. The orange triangle is the maximum of the one-dimensional marginalised pdf's.}
         \label{fits2d}
   \end{figure*}

\begin{table*}
\caption{Results of the fits for the RAVE data.}             
\label{t:fits}      
\centering          
    \tabcolsep 3.pt
\begin{tabular}{l lccccccccc }   
\hline\hline       
\multicolumn{2}{l}{Model}&$\pb$(\scriptsize{MAX})&E($\pb$)&$\sigma_{\pb}$&$\Ob/\Oo$(\scriptsize{MAX})&E($\Ob/\Oo$)&$\sigma_{\Ob/\Oo}$&$\rho_{\pb\Ob}$&E($\Ob/\Oo|\pb=30\deg$)&E($\Ob|\pb=30\deg$)\\ 
\multicolumn{2}{l}{}&($\deg$)&($\deg$)&($\deg$)&&&&&&($\kmskpc$)\\ 
\hline      
1& standard                &  45.&    48.&    17.&     1.89&     1.91&     0.08&     0.98&     1.83$\pm$     0.02&    54.0$\pm$     0.5\\
2& $\beta=0.$              &  65.&    48.&    17.&     1.97&     1.90&     0.07&     0.99&     1.83$\pm$     0.01&    54.0$\pm$     0.3\\
3& $\Vs=5\kms$             &  59.&    47.&    18.&     1.90&     1.85&     0.08&     0.98&     1.78$\pm$     0.02&    52.6$\pm$     0.5\\
4& free $\vo$              &  45.&    48.&    17.&     1.89&     1.91&     0.08&     0.97&     1.83$\pm$     0.02&    54.1$\pm$     0.6\\
5& $e_{\vr,\vphi}<15\kms$  &  41.&    48.&    17.&     1.89&     1.92&     0.08&     0.97&     1.84$\pm$     0.02&    54.4$\pm$     0.5\\
6& Binney dist.            &  44.&    45.&    19.&     1.89&     1.90&     0.09&     0.98&     1.83$\pm$     0.02&    54.1$\pm$     0.5\\
7& overest. dist. $30\%$   &  33.&    34.&    22.&     1.86&     1.87&     0.10&     0.99&     1.85$\pm$     0.02&    54.7$\pm$     0.5\\
8& underest. dist. $-30\%$ &  50.&    48.&    18.&     1.92&     1.92&     0.08&     0.95&     1.84$\pm$     0.03&    54.3$\pm$     0.8\\
\hline     
\end{tabular}
\end{table*}

Figure \ref{fits2d} shows the two-dimensional marginalised pdf's $\pb$-$\Ob$ (left), $\beta$-$\Ob$ (middle) and $\pb$-$\beta$ (right). The first panel presents a well defined peak. By contrast, the other panels show flatter distributions, especially for the slope $\beta$ of the rotation curve for which we do not obtain any constraint. In Table~\ref{t:fitssim} (Model 1) we give the details of the pdf of $\pb-\Ob$, that is the maximum of the probability, the expectation of each parameter E$(\pb,\Ob)$, the standard deviations of the distribution $\sigma_{\pb}$ and $\sigma_{\Ob/\Oo}$, and the correlation $\rho_{\pb\Ob}$. From Fig.~\ref{fits2d} we can observe (as in the simulations of Sect.~\ref{results0}) the strong correlation between $\pb$ and $\Ob$ with a correlation coefficient of $\rho_{\pb\Ob}=0.98$. Correlations between other parameters are much smaller: $\rho_{\beta\Ob}=-0.03$ and $\rho_{\pb\beta}=0.05$.

The maximum of the pdf is the orange cross in Fig.~\ref{fits2d} and is located at $(\pb, \Ob/\Oo)=(44.5\deg,1.89)$. 
The expected values are shown as a orange asterisk. For the choice of the parameters $\vo=238\kms$ and $\Ro=8.05\kpc$, the pattern speed of $1.89\Oo$ corresponds to $\Ob=56.0\kmskpc$. There are no significant differences between the maximum and the mean of the pdf for $\Ob$, as they differ only by $1\%$. For the bar's orientation $\pb$ we obtain a broader likelihood distribution than for $\Ob$. In the left panel of Fig.~\ref{fits2d} we see that the $1\sigma$ region (dotted red line) covers almost the whole range of $\pb$ (from $\sim 20\deg$ to $\sim80\deg$). The maximum of the pdf and its mean differ by $6\%$. The fit given by the maximum of the two-dimensional pdf is plotted on top of the data points in Fig.~\ref{vRdata} (black curves) labelled as ``2d max'', for three different values of $\beta$.

The fact that we can constrain the value of the pattern speed but not the orientation is mainly due to the small range of radius ($\sim600\pc$), in comparison to the toy model ($2-3\kpc$). This is expected from inspection of Fig.~\ref{theoriccurves}, where we see that different pattern speeds occupy distinct regions, while curves for different angles or slopes of the rotation curve can be rather close to each other for certain $R$ and may become indistinguishable given the errors. We also performed a test reducing the range of radii for our simulations to establish whether the quality of the constraint obtained depended on the radial range considered. In practise, for each of the four bands ($20$, $40$, $60$ and $80\deg$) we repeated the fit using only the innermost bins in a $600\pc$ radius range, and separately, the outermost bins. These tests showed that the effect of reducing the radius range doubles the uncertainty in the constrained $\pb$, making it of the order of $20\deg$, as for the RAVE data.

Although the errors could also be a cause of our weak constraint in the bar's orientation for the RAVE data, the precision to which $\volr$ is determined is not so different for the simulations and the data. For RAVE most of the errors are between $2$ and 3 $\kms$. For the simulations, although the uncertainties in the determination of $\volr$ are larger than $3\kms$ for $40\%$ of the bins (for the outer bins) we still determine the bar orientation more accurately in the simulations (to $10\deg$) than for the RAVE data ($\sim20\deg$).

One could marginalise the pdf's of Fig~\ref{fits2d} to obtain the best estimates for each individual parameter. If we proceed in this way, we get the maxima of the individual marginalisations as shown with a orange triangle in Fig~\ref{fits2d} left. For $\Ob$ the maximum of the two-dimensional pdf is similar to the one-dimensional maximum, differing only by $4\%$. However, note that for $\pb$ the maximum of the one-dimensional pdf is quite different ($55\%$) from that obtained from the two-dimensional $\pb-\Ob$ panel. This is analogous to what happened in the case of 
our simulations in  Sect.~\ref{test}, and is due to the global pdf being degenerate
and skewed especially in the $\pb$ direction. The resulting fit of the one-dimensional pdf's is the red curve shown in Fig.~\ref{vRdata} labelled as ``1d max''. This curve fits very poorly our data, showing once more that the one-dimensional pdf's do not capture the main correlation between the parameters and give misleading results.

Because of the tight correlation between $\Ob$ and $\pb$ and the large dispersion in the probability for $\pb$, we actually obtain a better fit and a tighter constraint when we use the appropriate combination of parameters. In the same manner as in Sect.~\ref{test}, under the bivariate normal approximation, using Eq.~\ref{e:linear} and Eq.~\ref{e:variance} we can establish a linear relation between $\pb$ and $\Ob$ that allows us to obtain the best estimate of $\Ob$ given a particular value of $\pb$. We obtain:
\beq\label{e:linearobs}
E(\Ob/\Oo|\pb = {\pb}_1) =1.91 + 0.0044\left({\pb}_1(\deg)-48\right)
\eeq
with standard deviation of $0.02$. In units of $\kmskpc$, when we fix the solar parameters to $\vo=238\kms$ and $\Ro=8.05\kpc$, this is:
\beq
E(\Ob|\pb = {\pb}_1) = 56.3+ 0.1316\left({\pb}_1(\deg)-48\right)
\eeq
with standard deviation of $0.5\kmskpc$. The green line in the left panel of Fig.\ref{fits2d} indicates this linear relation. An example is shown in the last two columns of Table~\ref{t:fitssim} (in units of $\Oo$ and in $\kmskpc$). For the angle of $\pb=30\deg$ we obtain a pattern speed of $54.0\pm0.5\kmskpc$. This model is shown on top of the data points in Fig.~\ref{vRdata} (blue curves) labelled ``30 $\deg$'' for three different values of $\beta$. We see that these curves fit better the data points, compared to the curve for the maxima of the one-dimensional marginalised pdf's (black curve). In the range of $10-45\deg$, which as explained in Sect.~\ref{intro} encompasses the bar orientations independently estimated in the literature, we would obtain a range of pattern speed of $51.3-55.9\kmskpc$.

\begin{figure}
   \centering
\includegraphics[width=0.4\textwidth]{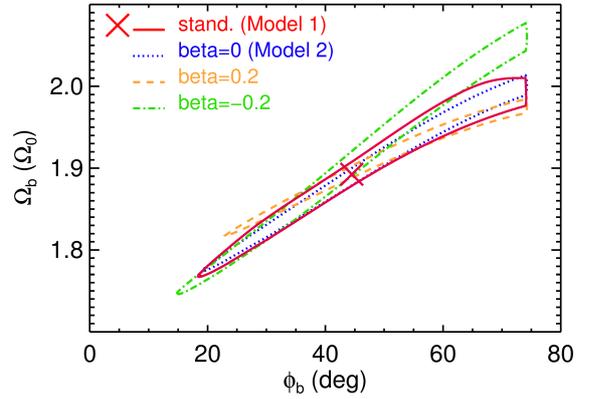}
  \caption{Two-dimensional likelihood in the $\pb$-$\Ob$ space marginalised over $\beta$ (red) and for slices of the three-dimensional probability at different values of $\beta$. The different lines show the $1\sigma$ confidence regions.}
         \label{fits2dbeta}
   \end{figure}

Figure \ref{fits2dbeta} shows the $1\sigma$ confidence limits for slices of the three-dimensional probability at different values of $\beta$ compared to Model 1 (marginalised over $\beta$). The curves do not differ significantly, meaning that the dependence on the $\beta$ parameter is not strong. For instance, when we fix our model to $\beta=0$ (Model 2 in Table \ref{t:fits})\footnote{In principle, we do not expect our Galaxy to have a rotation curve similar to the power laws in the model with a single $\vo$ and $\beta$ and that is the reason why in Model 1 we marginalised over $\beta$. However, recent studies point to a rather flat rotation curve. For instance, \citet{Honma12} using observations of masers claim $\beta=0.022\pm0.029$ (their parameter $\alpha$).}, 
we obtain similar results for the pdf of $\pb$ and $\Omega_b$ when compared to Model 1 (blue and red curves in Fig.~\ref{fits2dbeta}). For $\beta\ne0$ the curves are similar only for $\pb$ around the two-dimensional maxima (orange cross). This is why the fit obtained with the two-dimensional maximum (marginalised over $\beta$) fits the trend in Fig.~\ref{vRdata} for different values of $\beta$. For other angles (away from the 2D maximum), there
is though a slight dependence on $\beta$. 
Under the assumption that the parametrisation of the
rotation curve of Eq.~\ref{e:rotcurve} is valid, this figure demonstrates that $\pb = 30\deg$, 
and $\beta = 0.2$ (or in general positive $\beta$) are favoured by the data only if the pattern speed is higher. As clearly shown in Fig.~\ref{vRdata} with the blue-dashed curve, the linear relation of Eq.~\ref{e:linearobs} is not valid for positive $\beta$.

%%%%%%%%%%%%%%%%%%%%%%%%%%%%%%%%%%%%%%%%%%%%%%%%%%%
\subsection{Varying the Solar parameters}\label{results2}

 As explained before, we had to assume values for $\vo$, $\Vs$, $\Us$ and $\Ro$ to compute the individual $\vphi$, $\vr$ and $R$ from the observables. The adoption of this specific set of values has an effect on our derived bar's parameters. In this Sect., though, we shall see that the effects are in practise little or they lead to only a scaling on the obtained pattern speed.

First, we changed the value of $\Vs$ from $12\kms$ to $5\kms$ \citep[e.g.][]{Dehnen98b}.
To do this, we would have to recompute the velocities of our RAVE data points using this new value and re-run our method to find $\volr$ as a function of $R$. In particular, $v_\phi$ (which is one of the required observables) is obtained by adding to the ``measured'' heliocentric velocities the adopted values of $\vo$ and $\Vs$ and rotating them by an angle that depends on the position of the star in the disc (which in turn depends on the positions in the sky, distances from the Sun and the adopted value of $\Ro$). As a short-cut to this, one can see that for a particular angular band the change on the individual $v_\phi$ of a star due to a change $\Vs$ and $\vo$ will translate into a shift of the measured $\volr$.  Therefore, our new determinations $\volr$ can be recomputed from the old ones by adding a factor $(5-12)\cos(6\deg)\kms$, where $6\deg$ is the angle of the selected band. Model 3 of Table \ref{t:fits} shows the results for this change in $\Vs$. We see that this change slightly reduces the expected value of $\Ob$ to $1.85\Oo$ ($\Ob=54.7\kmskpc$). Also the conditional value of $\Ob$ for $\pb=30\deg$ is reduced to $52.6\pm0.5\kmskpc$. %The corresponding probability curve for $\Ob$ is superposed in Fig.~\ref{fits1d} (left, black vertical line). %When we use the prior in the bar's orientation, the best fit model (Model 5) is  $\Ob=1.82\Oo$ (dashed orange line in Fig.~\ref{fits1d} left).
On the other hand, for the band considered here (only at $6\deg$ from the line GC-Sun) the parameter $U_\odot$ has little influence on the computation of $\vphi$.

\begin{figure}
   \centering
 \includegraphics[width=0.4\textwidth]{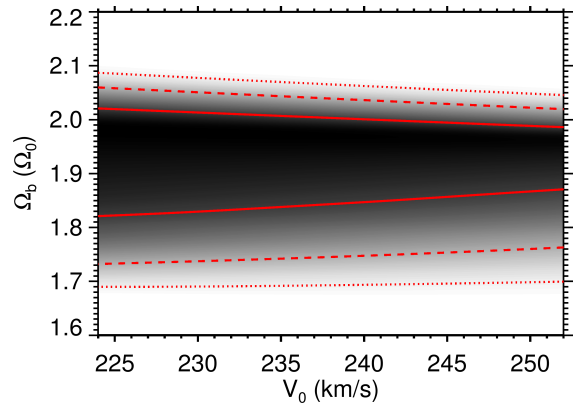}

  \caption{Two-dimensional (marginalised over the other parameters) likelihood function in the $\vo$-$\Ob$ plane for the RAVE data.% (top) and marginalised likelihood function for $\vo$ (bottom).}% Dotted, dashed and solid lines show the $1\sigma$, $2\sigma$ and $3\sigma$ confidence regions, respectively.
}
         \label{fits2d2}
   \end{figure}

Following the idea of the previous paragraph, we can also turn parameters such as $\vo$ into free parameters of the model without first having to compute the new velocities $\vphi$ of the data points for many values of $\vo$, then derive for each case the new determinations of $\volr$ and finally do the fitting process.
Model 4 in Table \ref{t:fits} is the best fit obtained when $\vo$ is a free parameter. We explore this parameter in the range of $[224,252]\kms$ \citep{Honma12} using steps of $1\kms$. This change, however, does not affect the determinations of $\Ob/\Oo$ and $\pb$ with respect to Model 1. Figure \ref{fits2d2} shows the pdf in the $\Ob$-$\Vs$ plane, with a correlation coefficient of $\sigma_{\Ob\Vs}=-0.03$. For $\vo$ of 224 or $252\kms$ we get similar best fit pattern speeds in the combined $\pb-\Ob$ pdf ($1.88\Oo$ and $1.90\Oo$, respectively). However, once scaled to the respective $\Oo$, the pattern speeds become $52.4$ and $59.5\kmskpc$. On the other hand, the pdf for $\vo$ is very flat as can be seen in Fig.~\ref{fits2d2}.

%%%%%%%%%%%%%%%%%%%%%%%%%%%%%%%%%%%%%%%%%%%%%%%%%%%
\subsection{Analysis of errors and biases}\label{results12}
\begin{figure}
   \centering
\includegraphics[width=0.45\textwidth]{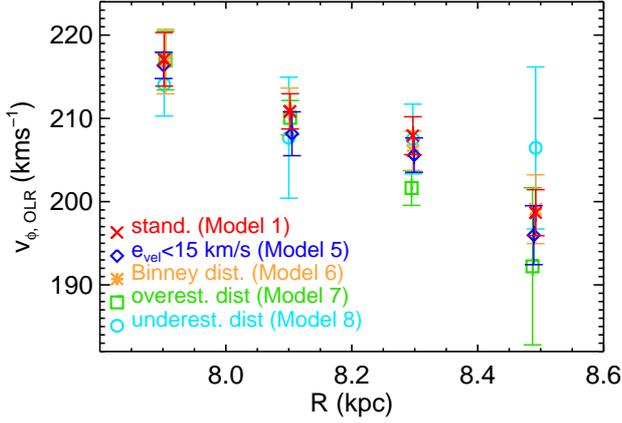}
  \caption{Position of the $\volr$ as a function of $R$ for the band at $\pb+6\deg$ for different cases of RAVE data.}
         \label{difdata}
   \end{figure}

\begin{figure}
   \centering
\includegraphics[width=0.4\textwidth]{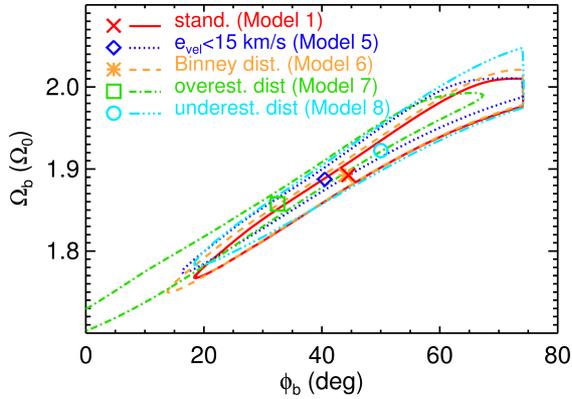}
  \caption{Two-dimensional (marginalised over $\beta$) likelihood in the $\pb$-$\Ob$ space for different cases of RAVE data. The different lines show the $1\sigma$ confidence regions and the different symbols are the two dimensional maxima of the likelihood. }
%The green square and triangle are the maximum likelihood values when the distances are decreased and increased by $30\%$, respectively.
         \label{fits2dcompare}
   \end{figure}

Here we explore the influence of the observational errors and biases on the analysis of Sect.~\ref{results1} by considering four possible cases. In the first case (Model 5) we consider only stars with velocity errors in the $\vr$ and $\vphi$ directions smaller than $15\kms$. This contains $35\%$ of our initial sample and 
has a total of 26076 stars. In the second case (Model 6), we use the distances from \citet{Binney13} obtained with the method 
by \citet{Burnett11}, instead of the one by \citet{Zwitter10}. Finally, we also explore how a bias in distance would affect our results. We redo the analysis considering the extreme cases of having distances overestimated (thus we reduce the original values) and underestimated (thus we increase the original values) by $30\%$. These are Models 7 and 8, respectively. The new measured values of $\volr$ for these four cases are shown in Fig.~\ref{difdata} with different symbols and colours. Using the same symbols, in Fig.~\ref{fits2dcompare} we show the maximum in the $\pb-\Ob$ plane, and the respective $1\sigma$ confidence limits are marked with a dotted line. In Table \ref{t:fits} we give the details of the two-dimensional pdf's in the same manner as the previous cases. The data points of each case are tabulated in Appendix \ref{tables}.

The results for these four additional cases are similar to Model 1. For example, for the distance method by \citet{Burnett11} (Model 6) we find an almost identical two-dimensional maximum and $1\sigma$ contour. For the four cases, the two-dimensional maxima are all located inside or very close to the $1\sigma$ contour of our standard Model 1. Moreover, the maxima are only shifted along the direction of degeneracy of Model 1 and their confidence regions also follow the same degeneracy. We do see, however, that the bar's orientation is more sensitive to observational errors and biases. 
For instance, the sample with the smaller errors (Model 5, blue diamond) has a maximum for a bar orientation that is $4\deg$ smaller than for Model 1 but the same expectation value. We also find a smaller (larger) bar's orientation in Model 7 and 8 when we correct the distances supposing that they were overestimated (underestimated), although they are consistent within the errors. These differences are because biases in distance systematically change the slope of the relation between  $\volr$ and $R$. We obtain similar results for the best pattern speed, whereas the value for a fixed orientation of $30\deg$ changes at most by $0.7\kmskpc$.

Finally, we also perform a test to assess whether the observational errors and the resulting blurring of the substructures could produce a bias in the derived models parameters. To this end we convolve the simulations of Sect.~\ref{test} with typical RAVE errors. More details are given in Appendix \ref{Aconv}.

The results of this test indicate that, as expected, the addition of errors produces a blurring of the structures which makes impossible the Hercules detection in certain radial bins, especially those located farthest from the simulated Sun. In the bins where Hercules can still be detected we observe a slight tendency to obtain smaller estimates of $\volr$ for the farthest bins and that appears to increase with distance. This seems to be due to contamination from stars located originally farther away and which fall in our sample because of their large distance error more than to kinematic errors only. In our parameter fits, the mean or expectation of both the pattern speed and the bar's orientation are consistent with the true parameters within $1\sigma$ in most of the cases. However, the obtained correlation between the angle and the pattern speed presents a small bias. In particular, if the true bar orientation is assumed, it leads to an overestimation of the pattern speed by an amount between $0.6$ and $1.1\kmskpc$ depending on the orientation.

Note that this test is rather simple at least regarding our RAVE error model and selection function and that it is based on a very idealized model. Moreover, the mentioned bias could be different depending on the model parameter region that we are probing. For these reasons one cannot conclude that our obtained bar parameters with RAVE in the case of the linear relation are overestimated by the cited amount. We rather use these results to estimate that the approximate systematic error due to the RAVE data precision in the determination of the pattern speed could be around $1\kmskpc$.

%%______________________________________________________________

\section{Discussion and conclusions}\label{conclusions}

We have derived the pattern speed of the Galactic bar from the analysis of the kinematics of the Hercules stream at different Galactocentric radii, assuming that Hercules is caused by the effects of the bar's OLR. The crucial observable
for this measurement is the azimuthal velocity of the saddle point that separates Hercules from the main part of the velocity distribution.

In particular, starting from the model by D00, we have derived an analytical expression for how the azimuthal velocity of the saddle point changes as a function of position in the Galaxy and its dependence on the properties of a barred potential, namely, the bar's pattern speed, orientation, and the slope and normalisation of the rotation curve. We then used data from the RAVE survey to measure 
this velocity as a function of Galactocentric radius. We have found that it decreases with radius
in a manner that is consistent with our analytic model. By fitting the measured trend, we have derived
the best fit parameters of the Galactic bar. To our knowledge, this is the first time that the information on how a moving group changes as a function of radius is used in deriving the parameters of the non-axisymmetries of the disc.

We tested the reliability of our analysis by comparing the model predictions with the ``measurements'' of the velocity of the saddle point in a toy model consisting of a test particle simulation. Although the analytical model was derived using the stellar orbital frequencies for simple power-law Galactic potentials, it was found to reproduce well the trends found with a more complex Galactic potential (with three components: halo, bulge and disc). Our method to locate the velocity of the saddle point successfully finds velocities that are consistent with the predictions and we recover the input parameters of our simulation in most cases inside the $1\sigma$ confidence region. We emphasise that a much more accurate constraint is obtained when the proper combination of $\Ob$ and $\pb$ (which are largely degenerate) and some prior information on $\pb$ are used.
%The recovered parameters present some offset with respect to the input parameters. However, these offsets are particular for the simulations presented here and difficult to quantify in a general way that could be included in the error bars of our study with RAVE data. Therefore, one should keep in mind that additional systematic errors may exist. However, w
%{\bf I am not convinced this should go here, in the sense that it could be part of a discussion, rather than conclusions, and it isn't totall relevant as a discussion of the results, it is mostly a limitation of the simulation... Perhaps some if this should go to the relevant section? If it is not there already...}}

Our model has provided new constraints for the parameters of the Milky Way bar. The likelihood function of the pattern speed and the bar's angle is highly degenerate. We find that the combined likelihood is maximum for a bar's pattern speed of $\Ob=(1.89 \pm 0.08) \times \Oo$, where the latter is the local circular frequency. Assuming a Solar radius of $8.05\kpc$ and a local circular velocity of $238 \kms$, this corresponds to a pattern speed of $56\kmskpc$ with a standard deviation of $\sim2 \kmskpc$.
%\textcolor{red}{However, as mentioned before {\bf here or in the main paper? If the latter then drop the ``as mentioned before''}, additional systematics could make the reported errors larger.} 
Also, because of the high correlation between $\pb$ and $\Ob$, we find that a better description of our best fit results is given by the linear relation $E(\Ob/\Oo|\pb = {\pb}_1) =1.91+ 0.0044\left({\pb}_1(\deg)-48\right)$ with standard deviation of $0.02$. For the angle of $\pb=30\deg$ we obtain a pattern speed of $54.0\pm0.5\kmskpc$, reducing further the uncertainty in this determination. In the range of bar's orientation of $10-45\deg$, as other independent studies suggest, we obtain a range of pattern speed of $51.3-55.9\kmskpc$. Tests made by adding typical RAVE errors to the the toy model indicate that RAVE-like errors could produce systematic errors in the pattern speed of around $1\kmskpc$ when estimated using this linear relation.

% When we use the prior of a bar's orientation being between $10\deg$ and $45\deg$, as other independent studies suggest, our best fit value for the pattern speed decreases to $1.87^{+0.02}_{-0.07}$ or $55.4^{+0.6}_{-2.1}\kmskpc$, reducing further the uncertainty in this determination. 
The determination of $\Ob$ in units of $\Oo$ is not very sensitive (typically only by a few centesimal digits) to the assumed Galactic parameters such as the circular velocity curve, the peculiar motion of the Sun, or to different distance determination methods or biases in distance. For instance, using a smaller value for the peculiar velocity of the Sun $\Vs$ reduces the best estimate for the pattern speed of the bar by $\sim 2\kmskpc$.

Our result agrees well with the pattern speed derived with other methods compiled in 
\citet{Gerhard11}. For instance, with a direct determination of the pattern speed using the Tremaine-Weinberg method a value of $2.15\pm0.54 \Oo$ was obtained \citep{Debattista02}. Several studies \citep[e.g.][]{Englmaier99,Fux99} that fitted hydrodynamical models to the observed CO $lv$-diagram give a combined value of $1.89\pm0.36\Oo$. Other determinations that are also consistent with ours come from the analysis of the kinematics of Solar neighbourhood stars. For example, the  observed trend with velocity dispersion of the Oort constant C could be due to the bar's effects with a pattern speed of $1.87\pm0.04\Oo$ according to \citet{Minchev07}. Also a value of $1.82\pm0.07\Oo$ could be responsible for the existence of several low-velocity streams in the Solar neighbourhood such as Pleiades and Coma Berenices, or Pleiades and Sirius \citep{Minchev10}. Compared to the value of D00 of $1.85\pm0.15\Oo$, who fitted the same stream as in the present work but only locally, we have obtained a consistent value but 
a tighter constraint. %chackrabarty

On the other hand, we find that the bar's orientation cannot be constrained well as the $1\sigma$ confidence region extends from $20$ to $80\deg$. This is mainly due to limitations in the data, in particular the small range of radii that it spans. This conclusion stems from the fact that we could constrain better this parameter for our toy model, which can be studied for a larger range of radii even if the the errors in the determination of the azimuthal velocity of the saddle point are comparable or larger than those of the data. Additional tests reducing the radius range of the simulations to make it similar to RAVE showed that the obtained uncertainties in the bar's orientation are larger and more comparable to the data.  %The parameters of the not on its orientation nor on the rotation curve. 

Alternative interpretations and models to explain the Hercules stream have been proposed in the literature. \citet{Fux01} in his study with test-particle and N-Body simulations suggested that near the external regions of the OLR the bar generates an overdensity of stars in velocity space that is made of chaotic orbits, 
%, that at certain bar inclination angles, gets a radial movement towards the outer disc. This is, therefore, 
which is a different interpretation to the scattering mechanism proposed by D00. \citet{Quillen11} showed that features similar to Hercules can be  %seen not only close to the OLR of the bar but also in different regions which can be
associated to the coupling of several spiral structures. Also \citet{Antoja09} found a group similar to Hercules in test-particle simulations containing only spiral arms, although a bar was required to move this feature to negative $U$. Our findings of the trend of Hercules with radius provide evidence that this stream may indeed be due to the effects of the bar's OLR, but whether these other models can account for such a trend remains to be seen.

An analysis with $i$) a sample covering larger range in radius and regions with different bar's orientation, $ii$) with more stars per bin, and $iii$) with smaller errors in distances and proper motions should allow us to constrain the bar's orientation and even better the pattern speed. The astrometric data from the ESA's Gaia mission will provide us with such numerous, extended and precise observations. A clear benefit would also be obtained when observations spanning 
several bands in azimuth could be used at the same time for tighter constraints.

\begin{acknowledgements}
We acknowledge funding support from the European Research Council under ERC-StG grant GALACTICA-240271. 
Funding for RAVE has been provided by: the Australian Astronomical Observatory; the Leibniz-Institut fuer Astrophysik Potsdam (AIP); the Australian National University; the Australian Research Council; the French National Research Agency; the German Research Foundation (SPP 1177 and SFB 881); the European Research Council (ERC-StG 240271 Galactica); the Istituto Nazionale di Astrofisica at Padova; The Johns Hopkins University; the National Science Foundation of the USA (AST-0908326); the W.M. Keck foundation; the Macquarie University; the Netherlands Research School for Astronomy; the Natural Sciences and Engineering Research Council of Canada; the Slovenian Research Agency; the Swiss National Science Foundation; the Science \& Technology Facilities Council of the UK; Opticon; Strasbourg Observatory; and the Universities of Groningen, Heidelberg, and Sydney. The RAVE web site is at http://www.rave-survey.org. 
 EKG acknowledge 
funding from SFB 881 ``The Milky Way System'' of the German 
Research Foundation, particularly through subproject A5. We also thank the anonymous referee for the helpful comments and suggestions which significantly improved the first version of this paper.

\end{acknowledgements}

\bibliographystyle{aa} % style aa.bst
%\bibliography{/data/users/antoja/bib/mybib} % your references Yourfile.bibQbÿ¿¿¿¿¿bÿ¿¿¿¿¿b
\bibliography{mybib}

\begin{appendix}

\section{$\volr$ and errors}\label{tables}

In this appendix we tabulate the data points of Figs. \ref{difdata} and \ref{vRdata} used for our different fits,  and their errors (Table \ref{tab:datapoints}).

\begin{table}[t]
\caption{Data points of Figs. \ref{difdata} and \ref{vRdata} together with their errors.}\label{tab:datapoints}
    \centering
   \begin{tabular}{lccc} 			\\[-2ex] 
     \hline\hline 	\\[-2.5ex]
       & $R $ 	& $\volr $ & $e_{\volr} $ 		\\
       & $(\kpc)$ 	& $(\kms)$ & $ (\kms)$ 		\\
      \hline		\\[-2.3ex]
standard               &7.902    &217.1    &3.2  \\ 
                       &8.102    &210.8    &2.1   \\
                       &8.297    &207.9    &2.3   \\ 
                       &8.492    &198.7    &2.8    \\\hline      
$e_{\vr,\vphi}<15\kms$ &7.901    &216.4    &1.6  \\  
                       &8.105    &208.2    &2.6  \\   
                       &8.299    &205.6    &2.1  \\   
                       &8.489    &196.0    &3.5   \\  \hline  
Binney dist.           &7.904    &216.8    &3.9 \\    
                       &8.101    &210.8    &2.8  \\  
                       &8.297    &206.1    &2.4   \\   
                       &8.492    &199.1    &4.1    \\  \hline      
overest. dist. $30\%$  &7.904    &216.9    &3.5  \\     
                       &8.102    &210.1    &2.1   \\   
                       &8.295    &201.6    &2.1  \\   
                       &8.487    &192.2    &9.4   \\   \hline     
underest. dist. $-30\%$&7.902    &214.1    &3.8    \\     
                       &8.100    &207.7    &7.3    \\     
                       &8.297    &207.5    &4.2    \\     
                       &8.492    &206.5    &9.7   \\     \hline

    \end{tabular}
\end{table}

\section{Effects of the RAVE errors}\label{Aconv}

 Here we perform some simple tests to assess whether the observational errors produce any bias in the derived models parameters. To this end we convolve the simulations of Sect.~\ref{test} with typical RAVE errors.
In particular, we use the median errors of the radial bins that we use in our analysis. That is we take a relative error in distance of $25\%$, error in proper motion (both in $\alpha$ and $\delta$) of $1.9$ mas/yr, and error in radial velocity of $0.9\kms$.

More in detail, we first orient the disc of the simulation so that the desired band with different position angles for the bar of $20, 40, 60$ or $80\deg$ has the same orientation with respect to the solar position ($6\deg$) as our RAVE band. We select particles in a band with the same angular width as in the observations and we then convolve the positions and velocities with the mentioned errors (assuming they are Gaussian). Afterwards we bin the simulation in the same way as the data. We select 4 bins located in the same (relative) position with respect to the simulated Sun as in the data to have particles with a similar distance distribution (and, therefore, similar errors in distance and in transverse velocity).

\begin{figure*}
   \centering
\includegraphics[height=0.5\textheight]{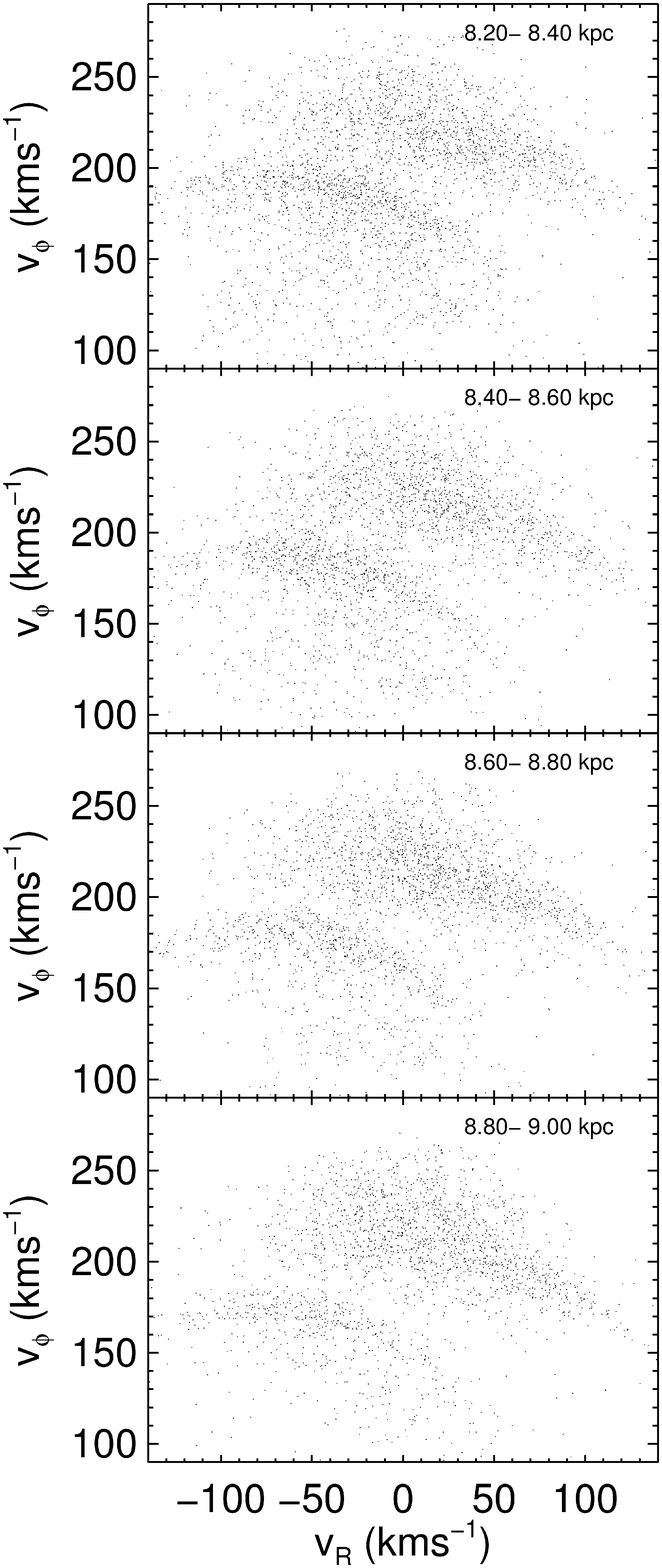} 
\includegraphics[height=0.5\textheight]{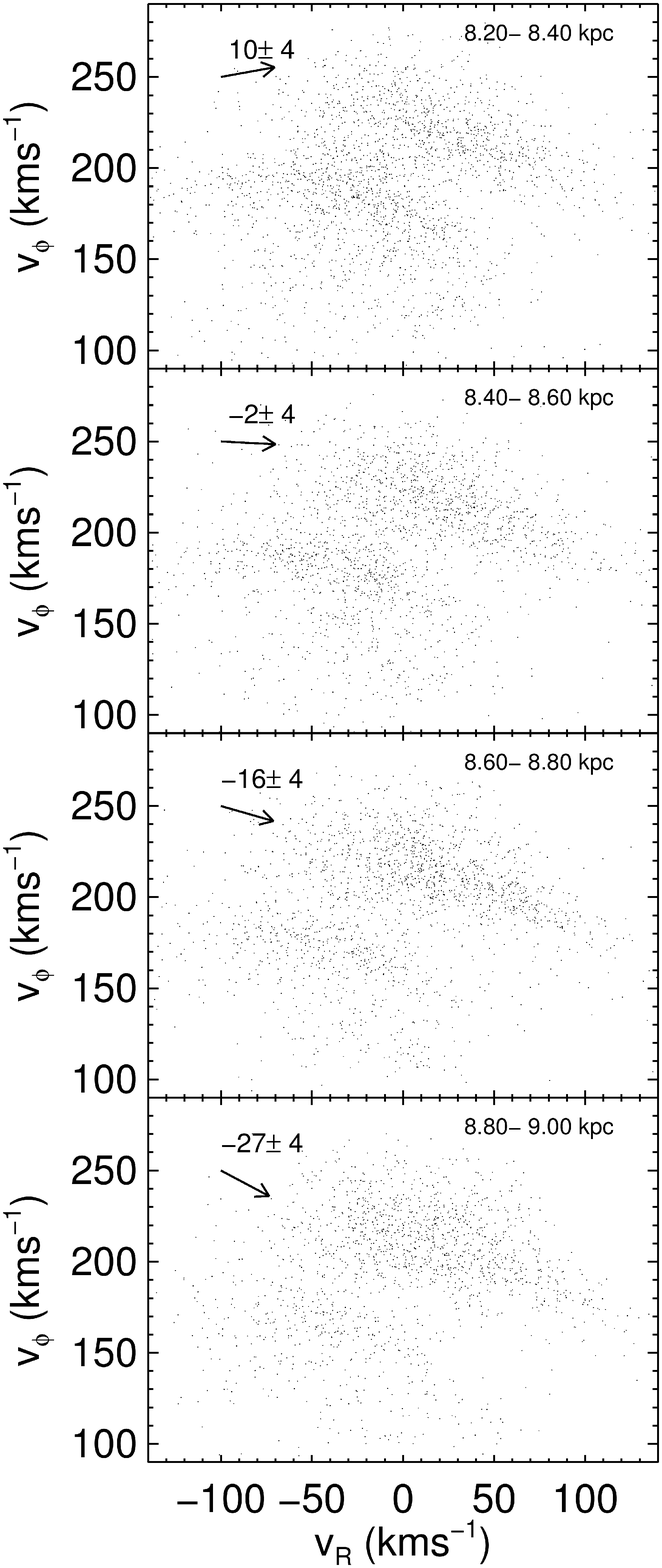} 
\includegraphics[height=0.5\textheight]{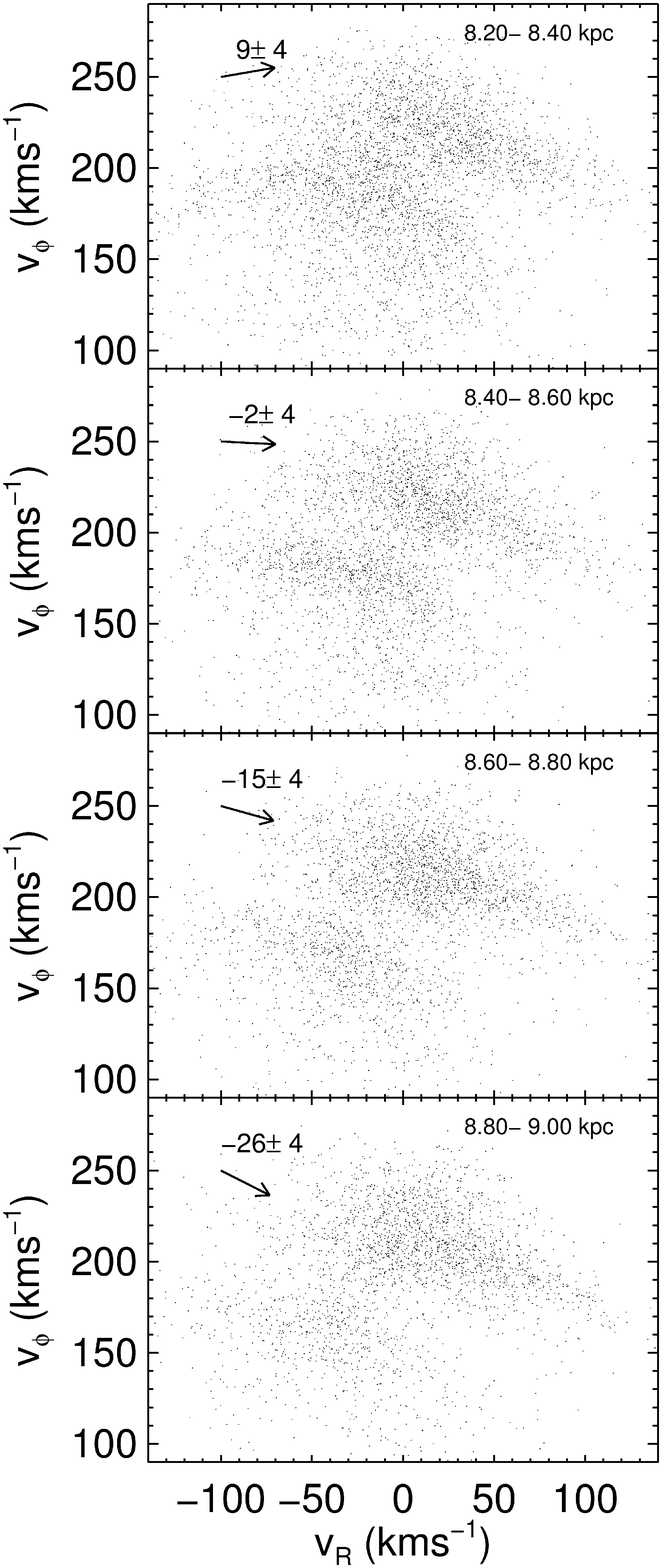} 
 \caption{Scatter plot of the velocities in bins in radius as indicated in the top right part of the panels for the band at $40\deg$ with no error convolution (left), with RAVE error convolution (middle) and with error convultion and allowing contamination from stars at different distances (right, see text). The numbers and the errors in the top left part of the panels are the median and dispersion of the direction of the transverse movement, taken as a counterclockwise angle with respect to the $v_R$ axis. The arrows indicate this median direction.}
         \label{conv}
   \end{figure*}

\begin{figure}
   \centering
\includegraphics[width=0.4\textwidth]{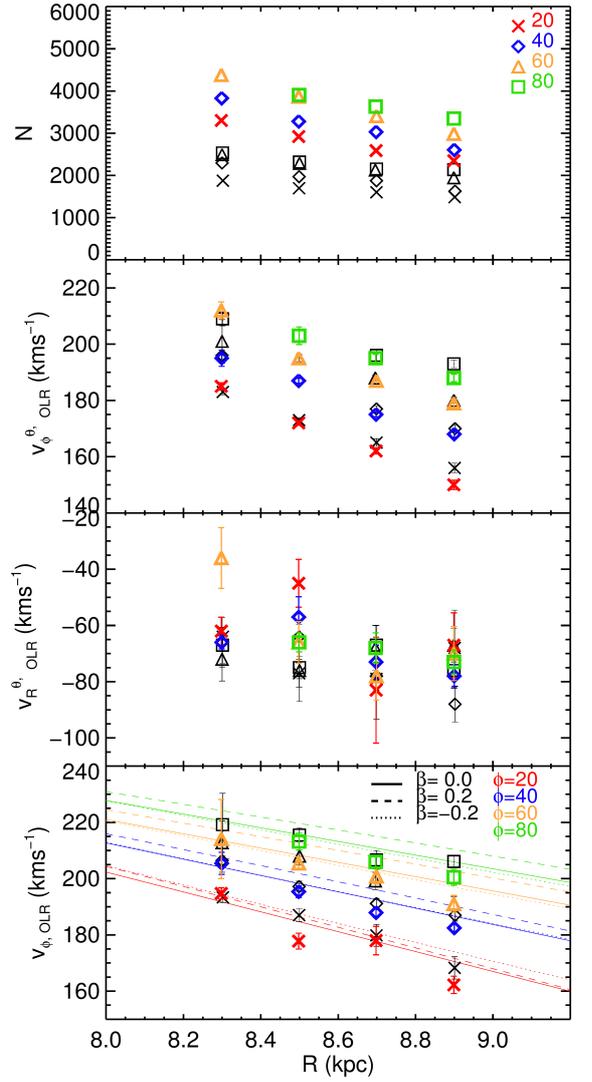} 
 \caption{Several measurements for the bands at different bar angles as in Fig.~\ref{vRsim} when we add RAVE errors (black symbols) and when we also allow for contamination from stars at other distances (colour symbols).}  
 \label{detconv}
   \end{figure}

An example of the final velocity distribution for the 4 bins for the band of $40\deg$ is shown in Fig.~\ref{conv} (middle). The new bins are smaller in extent and therefore have significantly less particles compared to Fig.~\ref{findvolrsim}. In order to do a proper comparison, we include here also the original simulation (i.e. without error convolution) in the same radial bins (left column). In the upper left corner we indicate the median direction of the transverse movement with an arrow. The blurring of the substructures occurs along this direction as this is the one influenced by errors in distance and proper motion, which are significantly larger than errors in line of sight velocity. The velocity distribution is distorted along this direction as explained in A12 and in fig. 9 of \citet{McMillan13}. For some of the bins (outside the range presented here), the blurring is large enough that the Hercules gap is no longer detected. The final determinations of $\volr$ for all bands (black symbols in Fig.~\ref{detconv}) fall close to the expected theoretical lines in most of the bins. Because of this no bias is observed in our final bar's parameter determination.

We now perform another test where we consider also particles originally located farther away in the disc but that, due to distance errors, end up in the selected band after error convolution. For this we first select particles with a maximum distance of $2.4\kpc$ from the Sun's position, we convolve with the RAVE errors, and finally we take the subset of particles in the bands. The limit of $2.4\kpc$ corresponds to the maximum distance at which a red clump star with absolute magnitude of $M_J=-0.87$ would be observed by the RAVE survey assuming that the upper magnitude limit of the survey is $J\sim11$. With this limit we avoid including particles in our band that were originally very far from the Sun and therefore, that would have never been observed because of the magnitude limit of the survey.

The example for the band of $40\deg$ is shown in the right column of Fig.~\ref{conv}. The final determinations of $\volr$ for all bands are the colour symbols in Fig.~\ref{detconv}. We observe in this case a slight tendency to obtain smaller estimates of $\volr$ for the bins that are far from the simulated Sun (the ones at larger radius) and that seems to increase with distance. However, as the RAVE bins are still quite close to the Sun with a maximum distance of $1.67\kpc$, this bias is not very significant.

The recovered parameters for all bands after RAVE error convolution are shown in  Table~\ref{t:fitssimerror}. We see how in most of the cases the correct model parameters both in angle and in pattern speed are recovered within the error bars. However, the recovered correlation between the angle and the pattern speed does present a bias. In particular, the recovered pattern speeds obtained with the linear relation using the correct (true) value of the bar's orientation are slightly larger than expected. In all cases the difference between the recovered value and the correct one is between 0.6 and $1.1\kmskpc$, increasing with bar's orientation. This is in all cases equal or less than $2\sigma$. We remark that we do not see this bias in the expectation values of the likelihood.

\begin{table*}
\caption{Results of the fits for the toy model when we add RAVE errors and contamination from stars at other distances. The input pattern speed is $\Ob=1.836\Oo=47.5\kmskpc$ for all cases. }             
\label{t:fitssimerror}      
\centering          
%\begin{tabular}{l l l l l l l l l ll }     % 11 columns 
    \tabcolsep 3.pt
\begin{tabular}{l lccccccccc }     % 11 columns 
\hline\hline       
\multicolumn{2}{l}{Input}&\multicolumn{1}{c}{$\pb$(MAX)}&\multicolumn{1}{c}{E($\pb$)}&\multicolumn{1}{c}{$\sigma_{\pb}$}&\multicolumn{1}{c}{$\Ob/\Oo$(MAX)}&\multicolumn{1}{c}{E($\Ob/\Oo$)}&\multicolumn{1}{c}{$\sigma_{\Ob/\Oo}$}&\multicolumn{1}{c}{$\rho_{\pb\Ob}$}&\multicolumn{1}{c}{E($\Ob/\Oo|\pb=input$)}&\multicolumn{1}{c}{E($\Ob|\pb=input$)}\\ 
\multicolumn{2}{l}{}&($\deg$)&($\deg$)&($\deg$)&&&&&&($\kmskpc$)\\ 
%&&&$(\kpc)$&$(\kms)$&$(\kms)$&&&$(\kmskpc)$&$(\deg)$\\ 
\hline      
 $20\deg$&& 0.&     7.&     8.&     1.77&     1.80&     0.04&     0.96&     1.86$\pm$     0.01&    48.1$\pm$     0.3\\
 $40\deg$&&30.&    37.&    21.&     1.82&     1.85&     0.10&     0.99&     1.87$\pm$     0.02&    48.3$\pm$     0.4\\
 $60\deg$&&66.&    40.&    23.&     1.89&     1.79&     0.09&     0.97&     1.87$\pm$     0.02&    48.3$\pm$     0.5\\
 $80\deg$&&80.&    43.&    22.&     1.88&     1.75&     0.08&     0.95&     1.88$\pm$     0.03&    48.6$\pm$     0.7\\
\hline                  
\end{tabular}
\end{table*}

\end{appendix}

\end{document}